\documentclass[journal]{IEEEtran}

\usepackage{graphicx}
\usepackage{amsmath,amssymb,amsfonts}
\usepackage{algpseudocode}
\usepackage{graphicx}
\usepackage{bbding}
\usepackage{pifont}
\usepackage{diagbox}
\usepackage{cite}
\usepackage{flushend, cuted}
\usepackage{lipsum}
\usepackage{stfloats}
\usepackage{cases}
\usepackage{multirow}
\usepackage{booktabs}
\usepackage{balance}
\usepackage{enumitem}
\usepackage{inconsolata}
\usepackage{bbm}
\usepackage{color}
\usepackage{bm}
\usepackage{hyperref}
\usepackage{threeparttable}
\usepackage{tabularx}
\usepackage{makecell}
\usepackage{booktabs} 
\usepackage{diagbox}
\usepackage[ruled,linesnumbered]{algorithm2e}

\ifCLASSINFOpdf

\else

\fi

\begin{document}

\title{{Data-Driven Privacy-Preserving Modeling and Frequency Regulation with Aggregated Electric Vehicles via Bilinear Hidden Markov Model}

\thanks{This work was supported by Natural Sciences and Engineering Research Council (NSERC) Discovery Grant, NSERC RGPIN-2022-03236, CRC-2023-00006,  and by Fonds de recherche du Quebec under Grant FRQ-NT 2023-NOVA-314338. \textit{(Corresponding author: Xiaozhe Wang.)}

The authors are with the Department of Electrical and Computer Engineering, McGill University, Montreal, QC H3A 0E9, Canada (e-mail: xiaozhe.wang2@mcgill.ca).
 }
}

% author names and affiliations
% use a multiple-column layout for up to three different
% affiliations
\author{Yiping Liu,~\IEEEmembership{Student Member,~IEEE}, Xiaozhe Wang,~\IEEEmembership{Senior Member,~IEEE}, Geza Joos,~\IEEEmembership{Life Fellow,~IEEE}
}

\maketitle

\begin{abstract}
Vehicle-to-Grid (V2G) technology allows bidirectional power flow for real-time grid support, making electric vehicles (EVs) well-suited for ancillary services such as frequency regulation. However, existing methods for flexibility estimation and coordinating aggregated EVs often rely on individual EV traveling information (e.g., arrival/departure time) and/or characteristic parameters (e.g., charging efficiency, battery capacity) as well as real-time state-of-charge (SOC),  %\color{red}should I add real-time SOC? \color{black}
%such as traveling time and initial SOC, or EV characteristic parameters like charging efficiency, rated charging power, and battery capacity, 
which raises privacy concerns and faces data quality issues. To address these challenges, this paper proposes a data-driven, privacy-preserving modeling and control framework for frequency regulation using aggregated EVs. The proposed method can provide accurate estimation for power outputs and flexibility of aggregated EVs and carry out effective frequency regulation without any individual EV information. %preserving user privacy and ensuring practical scalability. 
Simulation results %show that the proposed method can provide caparable power tracking and frequency requlation performance with model-based method 
validate the accuracy and effectiveness of the proposed method, which also outperforms the model-based %\color{blue} 
and federated learning-based method under SOC data inaccuracies. %when there are inaccuracies in SOC data. %systems bilinear hidden Markov model (bHMM) \color{red}Is bHMM the same as BLS? \color{black} for modeling aggregated electric vehicles (EVs), derived from a state space model (SSM).The bHMM effectively captures the dynamics of aggregated EVs across different states of charge (SOC), while representing state transitions with low real-time communication requirements. An expectation-maximization (EM) algorithm is employed to infer latent state transitions and predict power output trajectories. Furthermore, a data-driven modeling and control framework is developed to enable frequency regulation using aggregated EVs. Simulation results validate the accuracy and efficiency of the proposed method, even in the presence of missing SOC data and uncertain EV parameters. 
% Comparison studies are conducted to verify the effectiveness of the bHMM model for real-time prediction.
\end{abstract}


\IEEEpeerreviewmaketitle
\section{Introduction}
\IEEEPARstart{T}{he} 
increasing adoption of electric vehicles (EVs) is transforming the transportation and energy sectors. % By replacing petroleum with electricity, EVs offer a cleaner, more sustainable alternative to conventional vehicles. 
With Vehicle-to-Grid (V2G) technology, EVs act as mobile energy storage units capable of bidirectional power flow, enabling real-time charging or discharging based on grid demands. While individual EV capacity is limited, aggregated fleets—coordinated by EV aggregators (EVAs)—offer significant flexibility for ancillary services such as frequency regulation, load balancing, and peak shaving, thereby enhancing grid stability and supporting emissions reduction.

Integrating EVs into the grid via V2G represents a combination of data analytics and control theory. %Merging EVs into the grid is an efficient combination of data information and control theory. \color{black}
The existing EVA model and control generally fall into two categories: distributed and centralized. %can be classified into distributed and centralized schemes. 
%\color{blue}
In distributed control,  %mechanism, 
the control authority is delegated to individual EVs \cite{nimalsiri2019survey}, allowing them to make local decisions based on their own states and objectives. %\color{red} any other works?\color{black} 
In contrast, 
centralized schemes, such as the individual modeling method (IMM)\cite{zhang2016evaluation} and SOC-based sorting method \cite{meng2016dynamic}, %\color{red} how will these fit in the current version?\cite{zhang2016evaluation, wang2019state,wang2020electric}\color{black} 
rely on the EVA to collect all EV data and issue global control commands. While centralized control can achieve more optimal system-wide decisions, it becomes computationally heavy as the EV population grows \cite{song2018state}\color{black}.  %To realize real-time control, the individual modeling method (IMM)\cite{zhang2016evaluation}, SOC-based sorting method \cite{meng2016dynamic} 
% and \color{blue} Markov decision process model \cite{iversen2014optimal} \color{black}
%have been proposed, but it becomes computationally and communicationally heavy as the EV population grows \cite{song2018state}.
% optimization algorithms with different objects such as effective frequency regulation service\cite{kaur2017multiobjective} and maximum economic profit \cite{el2021optimization} \color{red} But are these modeling methods?\color{black}\cite{kaur2017multiobjective} % The individual modeling method (IMM) \cite{zhang2016evaluation}, which directly aggregates the power output and flexibility of each EV and often serves as a benchmark, has been validated in numerous studies \cite{liu2019optimal, lam2015capacity, wang2019state, wang2020electric} for its high accuracy in modeling and control.
%\color{red} mention IMM \color{black} 
%Centralized control can yield more optimal decisions under system-wide constraints, but it becomes computationally and communicationally heavy as the EV population grows \cite{song2018state}. 
%in a centralized scheme, the EVA collects all available information from each EV and dispatches control signals to determine their states. 
%Since the decision is made based on global constraints, it makes a better solution reducing optimization requirements. However, when dealing with a large number of EVs, the computation and communication complexity for data acquiring and transferring greatly increase \cite{song2018state}.
To improve scalability, studies \cite{wang2019state,wang2020electric,kiani2022extended} propose a state space %transition 
model (SSM) that classifies the large population of EVs into different  state- of-charge (SOC) state intervals %according to their state-of-charge (SOC) 
and controls the transition between different intervals to regulate frequency and energy imbalances. This approach keeps computational complexity dependent only on the number of intervals, not the EV population. %In this way, the computational complexity is not affected by the population of EVs and only depends on the number of state intervals. 
The extended state-space model (eSSM) %was further proposed in 
\cite{liu2025extended} further improves the prediction accuracy of power outputs and flexibility by accounting for %three special 
scenarios in which EVs temporarily lose regulation capability. %when  \color{red} eSSM\color{black}%The SSM model is then used to carry out frequency regulation. \color{red} Other model-based methods? \color{black}

%\color{blue}
% In addition to model-based method, learning-based methods \cite{tao2021data,forootani2023transfer,  ye2022learning, hao2023v2g, shi2023augmented,ding2020optimal,sadeghianpourhamami2019definition,yan2023real} have been proposed. 
%\color{blue} 
%\color{blue}
However, conventional model-based methods heavily rely on accurate individual EV information such as traveling data, characteristic parameters and SOC information, which are often unavailable or unreliable in practice due to %factors like 
battery degradation and measurement uncertainties \cite{tao2021data}. %\color{red} improve logic here\color{black} %To overcome these limitations, 
Learning-based methods\cite{tao2021data,hao2023v2g, shi2023augmented,ding2020optimal,sadeghianpourhamami2019definition,yan2023real} 
intended to address this limitation by operating with less individual EV information (see Table. \ref{tab:ev_info_requirements}). 
Nevertheless, both model-based \cite{zhang2016evaluation, wang2019state,wang2020electric, meng2016dynamic,liu2025extended,kiani2022extended}  and learning-based approaches\cite{tao2021data,hao2023v2g, shi2023augmented,yan2023real,ding2020optimal,sadeghianpourhamami2019definition}  methods %face challenges because they 
still rely on individual EV data—whether complete or partial, historical or real-time—as summarized in %Table I.either historical or real-time, complete or incomplete individual EV information %such as traveling information (e.g., initial SOC, arrival/departure time), characteristic parameters (e.g., charging efficiency, rated charging power, and battery capacity) and real-time SOC 
%as summarized in 
Table \ref{tab:ev_info_requirements}. %\color{red}should I add real-time SOC?
\color{black} %Relying on such information 
This reliance not only poses practical difficulties %, as the data is often incomplete or inaccurate 
due to sensor failures, communication delays, or data loss, but also raises significant privacy concerns. %but also poses practical difficulties, as the data is often incomplete or inaccurate due to sensor malfunctions, communication delays, or data sampling failures. 

To address privacy concerns, %Due to the privacy leakage concerns, 
%\color{blue} Due to the privacy leakage concerns, 
various techniques have been explored. %were applied in previous works. 
Differential Privacy (DP) methods has demonstrated effectiveness in real-time EV estimation and learning from EV databases. For example, \cite{ju2023local} designs a local DP-based data range query scheme to protect personal sensitive information on the EV side. Homomorphic encryption (HE), which allows computations on encrypted data without revealing its contents, % over any data without revealing the actual contents of that data, 
is also widely adopted. The study \cite{cheng2021homomorphic} proposes an HE-based collaborative distributed energy management system that enables joint optimization across multiple entities while safeguarding %preserving the privacy of 
sensitive operational data. %\color{blue}
However, HE and DP rely on data perturbation through noise addition or encryption, which inevitably reduces model accuracy. %are highly dependent on data tampering approaches (adding more noise and encryption), which inevitably reduces model accuracy. 
%Moreover, 
Balancing the privacy budget and model accuracy/learning performance remains a persistent challenge in practice.

Beyond these traditional privacy-preserving approaches, decentralized machine learning has emerged as %been implemented
% combined with DP and HE 
an effective solution for enhancing privacy. Unlike DP and HE, it avoids the need to transmit private individual EV data between the EVA and EVs  %to enhance privacy. Such approaches outperform DP and HE methods by avoiding 
%direct collection of private data 
by performing model training locally.  %directly, instead performing model training on local devices. 
Federated learning (FL) \cite{sani2022privacy} is the most prominent paradigm,  %is FL \cite{sani2022privacy}, 
which enables collaborative training while keeping individual EV data local. For instance, \cite{qian2023federated} develops a federated deep reinforcement learning algorithm to for optimal EV charging, %learn optimal EV charging strategies. 
and \cite{wang2025federated} combines DP and FL into an LDP-FL scheme, applying  %for autonomous EVs that uses 
local DP to protect sensitive data while mitigating performance loss from noise.   
% \cite{lee2024multilevel} develops a three-level DRL-based EV charging algorithm that incorporates DP into EV energy data processing.
% Apart from traditional privacy preserving notions, decentralized machine learning methods have been combined with DP and HE. The combined scheme does not collect private data from individuals and only carry out model training at local end devices. \cite{lee2024multilevel} develops a three-level deep reinforcement learning (DRL)-based EV charging algorithm by processing EV energy data with DP. The most prominent one is federated learning (FL). 
% % \cite{pokhrel2022data} 
% \cite{wang2025federated} presents a LDP-FL for autonomous EVs by using LDP to process private autonomous EV data and aims to minimize the effect of added noise with a performance loss constraint mechanism. 
%While FL enables collaborative training while preserving the privacy of local EV data, 
Although FL enhances privacy by avoiding data transmission, its local model training still depends on accurate individual EV data \cite{chellapandi2023federated}. Moreover, the global model's accuracy and convergence cannot be guaranteed, particularly when individual EV data contains errors 
%as the number of clientsyou mean EV? increases 
as will be demonstrated in Section V. %\color{red} TBD. How about communication burden?\color{black} \color{blue} In addition,  the communication burden is significantly higher.

%Despite these advantages, FL performance depends heavily on %of the global model remains highly dependent on 
%the quality of data from individual EVs~\cite{chellapandi2023federated} during local training. Though individual EV data is not transmitted between EVA and EV, the local training  through Its accuracy and convergence cannot be guaranteed, especially as the number of clients\color{red} you mean EV? \color{black} increases as will be discussed in Section. V of this paper.
% it requires each node possesses computation ability. With the number of clients increase, convergence can not be ensured and communication requirements also increase significantly \cite{yurdem2024federated}.
%, which provides privacy preserving learning as the centralized authority only send the model to learn from end users in a decentralized manner. 

% However, when higher privacy requires more noise, which can reduce data accuracy and undermines control results.
\color{black}

To address these challenges, this paper proposes a data-driven privacy-preserving modeling and control framework for frequency regulation in V2G. Building on the scalability of the eSSM structure, we reformulate it into a bilinear hidden Markov model (bHMM) to eliminate the dependence of model outputs (aggregated power output and flexibility) on the SOC of individual EVs, thereby eliminating the need for individual SOC measurements. The Expectation-Maximization (EM) algorithm, a maximum likelihood approach, is then employed to estimate the bHMM model parameters using only aggregated EV data (i.e., the total power output from all EVs). Based on the estimated model, a model predictive control (MPC)–based broadcast strategy is developed for real-time frequency regulation. The proposed approach decentralizes part of the control authority to individual EVs, allowing them to operate within their local constraints while collectively supporting grid frequency regulation. %enabling them to satisfy their local charging constraints while collectively supporting grid frequency regulation. 
To the best of our knowledge, this work seems to be the first to estimate aggregated EV flexibility and power output without requiring any individual EV information. %The key advantages of the proposed data-driven bHMM modeling and control framework are summarized as follows: %by decentralizing part of the control authority to individual EVs, ensuring their charging constraints are satisfied while supporting frequency regulation based on real-time grid demands. %, while obeying their charging constraints and support frequency regulation based on real-time grid demands. % The global control signal, designed without access to individual EV information, is broadcast to all EVs, enabling each vehicle to make local decisions consistent with its operational constraints while collectively achieving the desired aggregate power-tracking performance.
% Finally, a real-time model predictive control (MPC) strategy is developed to dispatch EV power while obeying their charging constraints and support frequency regulation based on real-time grid demands. 
%\color{blue}Finally, a real-time model predictive control (MPC) strategy is developed to dispatch EV power by decentralizing part of the control authority to individual EVs, ensuring their charging constraints are satisfied while supporting frequency regulation based on real-time grid demands. \color{black} 
%To the best of our knowledge, this work represents the first attempt of estimating the aggregated EV flexibility and power output and carrying out frequency regulation without any individual EV information. 
The key advantages of the proposed data-driven bHMM modeling and control framework are as follows:

\color{black}

% Frequency stability is one fundamental requirement in the operation of power systems, directly impacting the reliability and efficiency of the grid. When there is a sudden change in load, the synchronous generators' inertia respond to maintain a relatively stable speed in a short time. To stabilize the frequency, the frequency control system intervenes by adjusting the output power of the generators. This often involves using PID controllers or other regulating mechanisms to change the generator's power output and eliminate frequency deviations. Finally, the automatic generation controls bring the frequency back to the nominal value, which is called the secondary frequency regulation.\cite{kundur2004definition} \textcolor{blue}{//}
% When a sudden change in load 
%\color{blue}
% In light of the above discussion, we propose a power trajectory and flexibility prediction method for the aggregated EVs based on the state transition of aggregated EVs. The conventional SSM is transferred into an extended state space model (SSM) by adding two extra states.

%This paper proposes a data-driven modeling and control method of aggregated EVs. 
%The key advantages of the proposed data-driven bHMM modeling and control framework are summarized as follows:

1) Compared to existing model-based \cite{zhang2016evaluation, wang2019state,wang2020electric, meng2016dynamic,liu2025extended,kiani2022extended} and learning-based methods \cite{tao2021data,hao2023v2g, shi2023augmented,yan2023real,ding2020optimal,sadeghianpourhamami2019definition}, the proposed data-driven bHMM modeling method %preserves user privacy as it 
requires no individual EV information such as historical or real-time SOC, charging/discharging power and efficiency, battery capacity, arrival/departure time. In contrast to privacy-preserving federated learning methods\cite{qian2023federated, wang2025federated}, which still require collecting individual EV data for local training, the proposed method eliminates both the collection and transmission of such data. %that still require the collection of individual EV data for training local clients, the proposed method requires neither the collection nor the transmission of individual EV data.  

2) By integrating the structured eSSM with the EM learning algorithm, %model eSSM with advanced learning algorithm EM algorithm, 
the proposed method can accurately estimate the aggregated EV flexibility and power output, with guaranteed \textit{convergence}. Building on these estimates, an MPC-based broadcast strategy is developed for frequency regulation. %to achieve decentralized control, ensuring that each EV satisfies its operational constraints while collectively realizing the target aggregate response. %a model predictive control (MPC)–based broadcast strategy is developed for frequency regulation. 
The global control signal, designed without individual EV data, is broadcast to all EVs, enabling each EV to make decisions based on its mode and SOC, thereby satisfying its own operational constraints while collectively achieving the desired aggregate power-tracking performance. %satisfy its operational constraints while collectively achieving the desired aggregate power-tracking performance.
%Leveraging the estimated aggregated EV flexibility and power output, a model predictive control (MPC)-based broadcast strategy is developed for frequency regulation,   enabling each EV to respect its
%own operational constraints while collectively achieving the
%desired aggregate power-tracking performance. 

%The proposed data-driven bHMM modeling method preserves user privacy as it requires no individual EV information such as historical or real-time SOC, charging/discharging power and efficiency, battery capacity, arrival/departure time. 

%2) The proposed method, integrating sliding window with EM algorithm, requires only periodic model updates (every 3 mins), %\color{blue} with new aggregated EV information and guarantees convergence\color{black}, 
%which reduces real-time communication burden while providing accurate estimation for power outputs and flexibility of aggregated EVs.  

%\color{blue}
3) Simulation results demonstrate that the proposed modeling and control framework achieves power-tracking and frequency regulation performance comparable to the eSSM approach and superior to the FL-based method when individual EV SOC data are accurate, while requiring significantly less communication since no individual EV information is needed. Moreover, when SOC data are inaccurate, the proposed method outperforms both the model-based eSSM and FL-based approaches in power tracking and frequency regulation.
\color{black}

The rest of the paper is organized as follows. Section II introduces modeling methods of %the eSSM of 
aggregated EVs. %with different characteristics. 
Section III presents the proposed data-driven privacy-preserving modeling method based on the bHMM. %formulates a data-driven modeling and control framework based on the eSSM. 
Section IV describes the frequency regulation strategy and the complete modeling and control framework. %with the proposed framework. 
Section V presents and discusses numerical validation results. Section VI concludes the paper and outlines future work. 

\begin{table}[t]
\centering
\caption{Requirements for Individual EV Information of Different Control Methods$^*$}
\vspace{-0.2cm}
\label{tab:ev_info_requirements}
% \vspace{-0.5em}
\renewcommand{\arraystretch}{1.1}

\begin{threeparttable}
  % 只缩放表格，不缩放注释
  \resizebox{\linewidth}{!}{%
    \begin{tabular}{llccc}
    \hline\hline
    \textbf{Method Type} & \textbf{Approach} & \textbf{SOC Value} & \textbf{C P} & \textbf{T P} \\
    \midrule
    \multirow{3}{*}{Model-based} 
        & IMM~\cite{zhang2016evaluation} & \checkmark & \checkmark & \checkmark \\
        & SOC-based sorting~\cite{meng2016dynamic} & \checkmark & \checkmark & \checkmark \\
        & State Space Model~\cite{wang2019state, wang2020electric, liu2025extended,kiani2022extended} & every 3 mins & \checkmark & \checkmark\\
    \midrule
    \multirow{4}{*}{Learning-based} 
        & Modified GAN~\cite{tao2021data} & \XSolidBrush & \checkmark & \checkmark \\
        & Probabilistic Distribution~\cite{hao2023v2g, shi2023augmented} & \checkmark & historical& historical\\
        & Markov Decision Process~\cite{ding2020optimal, sadeghianpourhamami2019definition} & \XSolidBrush & historical & historical\\
        & Feedback-based Online Algorithm~\cite{yan2023real} & \XSolidBrush & \XSolidBrush & \checkmark \\
    \midrule
    \textbf{This work} & \textbf{bHMM-based MPC} & \XSolidBrush & \XSolidBrush & \XSolidBrush \\
    \hline\hline
    \end{tabular}%
  }
\end{threeparttable}

\begin{tablenotes}[flushleft]
\footnotesize
\item * \textbf{C P} denotes characteristic parameters, and \textbf{T P} denotes traveling parameters. \\
A \checkmark\ indicates that real-time data are required, while “historical” refers to \\ the use of past information rather than real-time measurements.
\end{tablenotes}
\vspace{-0.1cm}
\end{table}
% add a line of this paper

% transferred to a bilinear hidden Markov model (bHMM) and we implement the EM algorithm to determine the parameters. It works under the missing EVs' SOC states, characteristic parameters, and traveling parameters. 

%=======================Modeling=============================
\section{%\color{blue}
Preliminary: Modeling Methods for Aggregated EVs}
\vspace{-0.2cm}
In this section, we introduce the state-space modeling structure underlying our approach. %exploiting its framework without requiring individual EV data. %. It is worth noting that we only exploit the structure, but we will not require the model parameters and state variables that depend on individual EV data. 
We begin with the benchmark Individual Modeling Method (IMM)~\cite{zhang2016evaluation}, which describes the state and operational dynamics of a single EV. Because the IMM scales linearly with fleet size, studies~\cite{wang2019state, wang2020electric} proposed a state-space model (SSM) to enable scalable EV control with reduced communication and computation costs. Our previous work~\cite{liu2025extended} further extended this to the extended state-space model (eSSM), which accounts for scenarios where EVs temporarily lose regulation capability. However, the eSSM still requires detailed individual EV data to construct model matrices. In this work, we retain the eSSM structure but develop a data-driven approach that eliminates the need for individual EV data, thereby preserving user privacy. Model parameters are estimated solely from aggregated fleet-level information as will be elaborated in the next Section.
\vspace{-0.4cm}
\subsection{%\color{blue}
Individual Modeling Method of a Single EV}
\vspace{-0.1cm}
%\color{red} can the section title be changed to individual modeling method for a single EV? as this is what is used in simulation?\color{black}
%\color{blue}
%Considering the daily trip of an EV, each EV is assumed to be either in a driving state or connecting state. 
When an EV is connected to the grid, it operates in one of the three modes: % three types of connecting states are defined as follows:
(i) Charging Mode (CM): receiving power from the grid; %Active power flows from the grid to the EV; 
(ii) Idle Mode (IM): connected with no power exchange; %the EV connects to the grid with no active power exchange;
(iii) Discharge Mode (DM): supplying power %active power flows from the EV 
to the grid with rated discharging power. The SOC variation of an individual EV $i$ is described by (\ref{eq:SOC_variation}): %\color{red} check the third line \color{black}
\vspace{-0.1cm}
\begin{equation}
\vspace{-0.2cm}
    {\footnotesize
    \begin{aligned}
        S_{i}(k+1) = 
        \begin{cases}
        S_{i}(k) + P_{i}(k)\cdot \eta_{c,i} \cdot \Delta t /Q_{i}, & P_{i}(k) = P_{c,i} \\
        S_{i}(k), & P_{i}(k) = 0 \\
        S_{i}(k) - P_{i}(k)\cdot \Delta t/(\eta_{d,i} \cdot Q_{i}), & P_{i}(k) = P_{d,i}
        \end{cases}
    \end{aligned}
    }
    \label{eq:SOC_variation}
\end{equation}
%\color{red}$\eta$ in (1) need modification\color{black}
where $\Delta t$ is the time step; $S_{i}(k)$ is the SOC at the time $ k\Delta t $; $P_{i}(k)$ is the real-time power output; $Q_{i}$ is the battery capacity; $P_{c,i}/P_{d,i}$ is the rated charging/discharging power; $\eta_{c,i}/\eta_{d,i}$ is the charging/discharging efficiency.
To provide ancillary service, an EV can switch between different connecting states. %The transition between different connecting states is illustrated 

\begin{figure}
\vspace{-0.6em}
\centering
\includegraphics[width=0.76\columnwidth]{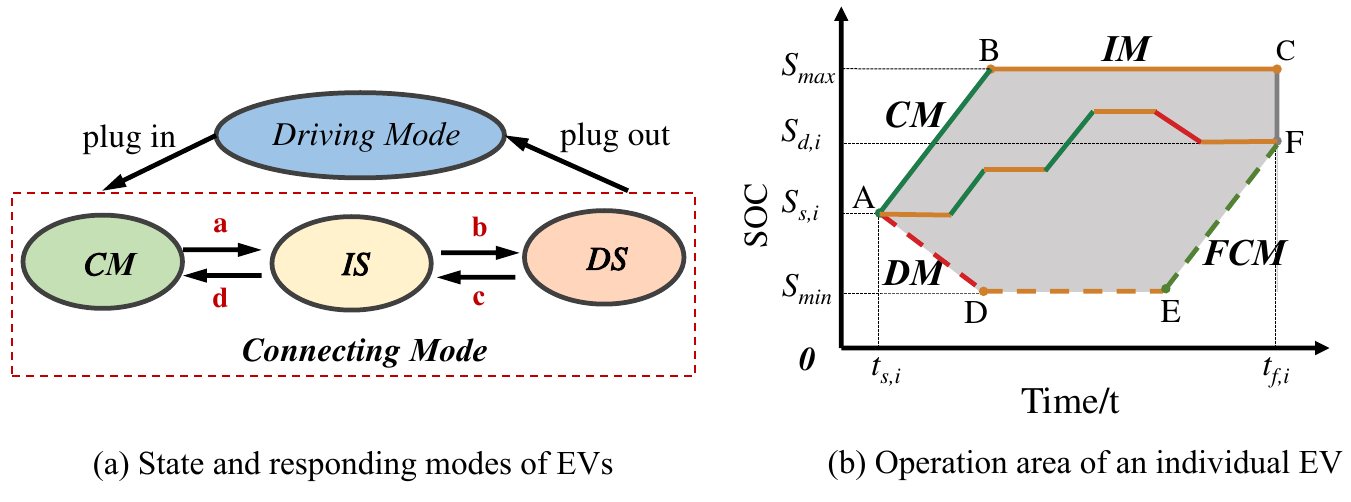}
\vspace{-0.3cm}
\caption{Individual charging model of an EV.}
\label{fig:states}
\vspace{-0.2cm}
\end{figure}

As shown in Fig. \ref{fig:states}. (a),  %\color{blue}%\color{red} cite properly \color{black}
four primary response modes are defined: a. `CM$\to${IM}'; b. `IM$\to$DM’; c. `DM$\to$IM'; d. `IM$\to$CM'. %These are the four responding modes of an EV. 
`CM$\to$DM’ (`DM'$\to$`CM') is assumed to be the combination of a and b (c and d). 
% `DS $\to$ CS' is the combination of c and d. \color{black}
% \begin{figure}
%     \centering
%     \includegraphics[width=0.67\linewidth]{Contents/figs/operation_area.pdf}
%     \caption{Operation area of an individual EV}
%     \label{fig:operation_area}
% \end{figure}
The operation area of an individual EV $i$ when connecting to the grid is illustrated in Fig. \ref{fig:states}. (b). $t_{s,i}$/$t_{f,i}$ is the arrival/departure time; $S_{\min}$/$S_{\max}$ is the minimum/maximum SOC value. 
$S_{s,i}$ is the initial SOC value when plugging in, and $S_{d,i}$ is the minimum demanded SOC when the EV finishes charging at $t_{f,i}$.
The upper bound `A-B-C' indicates that the EV enters CM immediately at $t_{s,i}$ and keeps CM until reaching $S_{max}$; the lower bound `A-D-E-F' indicates that the EV enters DM at $t_{s,i}$ and turns into IS until reaching $S_{\min}$. 
Specifically, to ensure that $S_{d,i}$ is achieved by $t_{f,i}$, the EV may enter a Forced Charging Mode (FCM), denoted as `E-F'. \color{black} Once connected at $t_{s,i}$, the EV can operate only within this region. %\color{red} Consider putting Fig. 1 and 2 side by side to save space. Also, did you explain FCS as it shows up in Fig. 2? \color{black}
% After an EV $i$ connects to the grid at $t_{s,i}$, it can only operate in the operation area. %\color{red} figure missing and cite properly \color{black}
\vspace{-1em}
\subsection{The Extended State Space Model of Aggregated EVs}
\vspace{-0.1cm}
To address the dimensionality challenge of individual EV models (\ref{eq:SOC_variation}), %The state space model (SSM) was proposed in \cite{wang2019state} to handle the dimensionality explosion of individual EV charging model (\ref{eq:SOC_variation}). As shown in Fig. \ref{fig:state_transition}, 
the SSM \cite{wang2019state}  discretizes the SOC range $[S_{\min}, S_{\max}]$  into $N$ intervals for each of the three connecting modes (CM, IM, and DM), resulting in $3N$ states labeled $1$ to $3N$. %The model also includes a forced charging state (FCS) to ensure the required SOC $S_{d,i}$ is reached by 
%$t_{f,i}$, during which the EV loses all flexibility. %the state of charge (SOC) variation range, denoted as $[S_{\min}, S_{\max}]$ is divided into $N$ discrete states. Since there are three different connecting states, the state intervals are labeled as $1, 2, \cdots, N, \cdots, 2N, \cdots, 3N$. Additionally, the %\color{blue} 
%\color{red}do you mean SSM model?\color{black}
%accounts for a scenario in which the electric vehicle (EV) enters a forced charging state (FCS) to ensure that the demanded SOC $S_{d,i}$ is reached by $t_{f,i}$. In this state, the EV loses both its charging and discharging flexibility. %\color{red} 
However, as shown in %[your conference paper],  (one sentence to explain the need for eSSM)\color{blue} 
\cite{liu2025extended}, the SSM may inaccurately estimate flexibility of aggregated EVs or respond to power requirements %when EVs reach full charge or discharge.  %\color{red} 
%SSM may fail to %....\color{blue} 
%estimate the flexibility of aggregated EVs or respond to power requirements by the grid %\color{red} because...\color{blue} 
because it overlooks the scenarios where EVs are fully charged and discharged. \color{black} 
To remedy this, %To create a more comprehensive framework, 
%we can 
an eSSM was proposed in \cite{liu2025extended}, incorporating three additional states: %our previous work \cite{liu2025extended} expands the state space model into an extended state space model (eSSM) by incorporating two additional scenarios: 
(i) IM with full discharge 
$S_{\min}$, (ii) IM with full charge $S_{\max}$, (iii) FCM. These are labeled as states 
$3N+1$, $3N+2$, and $3N+3$, respectively. EVs in these three states lose discharging, charging, or any regulation capacity.  %both regulation capacity, respectively. \color{red} tbd\color{black} %When the EV becomes fully charged and transitions to $S_{\max}$ in IS, it cannot revert to CS and lacks any capacity for charging regulation; (ii) If the EV becomes fully discharged, it also loses its discharging flexibility.
% In the state space model (SSM) shown as Fig. [], the SOC variation range $[S_{\min}, S_{\max}]$ is divided into $N_{s}$ states.  \color{blue}Besides, it considers the case that the EV may enter the force charging state (FCS) to ensure $S_{d,i}$ is attained at $t_{f,i}$, when it loses both charging and discharging flexibility. And it can be expanded into an extended state space model (eSSN) by considering other two cases: (i) When the EV gets fully charged and enters IS, it cannot be switched back to CS and has no charging regulation capacity; (ii) If the EV gets fully discharged, it loses its discharging flexibility.
%\color{red}FCS doesn't seem to be explained \color{black}  
%We use the intervals $3N+1$, $3N+2$, and $3N+3$ to describe the three cases respectively. 
The distribution of aggregated EVs is %can be 
denoted as a state vector $\bm{x}\in \mathbb{R}^{(3N+3)\times 1}$. $x_{n}$ (${1 \leq n \leq 3N}$) denotes the proportion of EVs in the state $n$, $x_{3N+1}$/$x_{3N+2}$ indicates the proportion of EVs in IS with $S_{\min}$/$S_{\max}$ %\color{red} to be consistent (i) and (ii) need to change order? \color{black}, 
and $x_{3N+3}$ represents those in FCM.
\begin{figure}
% \vspace{-0.3em}
    \centering
    \includegraphics[width=0.95\linewidth]{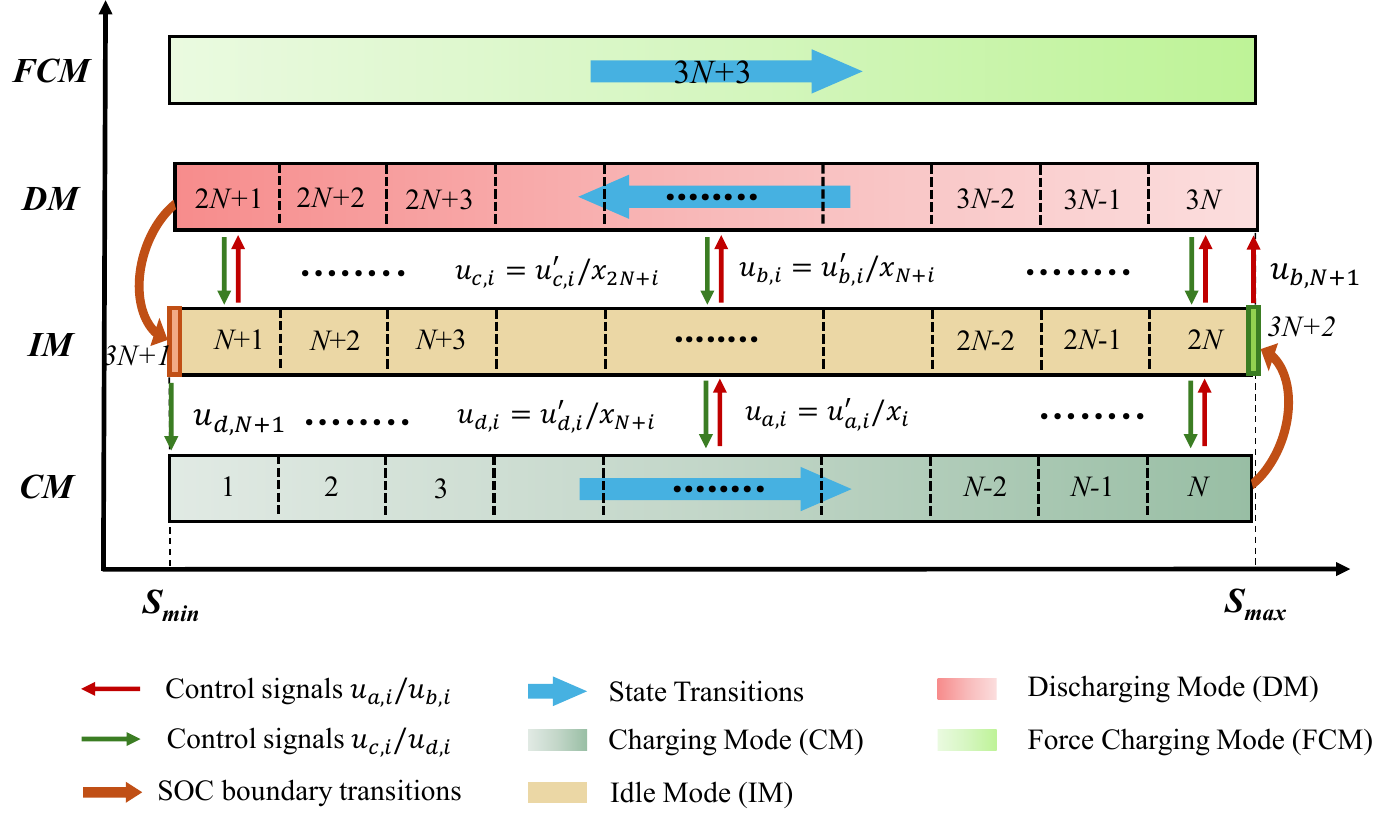}
    % \vspace{-0.2cm}
    \caption{State transition of aggregated EVs. Modified from Fig. 4. in \cite{wang2019state}.}
    \label{fig:state_transition}
    % \vspace{-0.1cm}
\end{figure}

Each EV can find its location in the state space according to its connecting states and SOC value. Blue arrows in Fig. \ref{fig:state_transition}  show transitions within the same connecting state, while brown arrows mark boundary transitions (`$x_{N}\rightarrow x_{3N+2}$' and `$x_{2N+1} \rightarrow x_{3N+1}$'). %\color{red} I don't understand why it is 2 \color{black} 
Assuming a uniform EV distribution across intervals, a Markov transition matrix $\bm{A}\in \mathbb{R}^{(3N+3)\times (3N+3)}$
  can be constructed to model state transitions, where each element $A_{m,n}$ is the probability of moving from state $n$ to $m$, %\color{red} or m to n? \color{black} 
  derived analytically from EV parameters \cite{wang2019state}.
%  The blue arrows in Fig. \ref{fig:state_transition} indicate the automatic state transition between states in the same connecting n_pn_pstate. The boundary transition (`$\bm{x}_{N}\rightarrow \bm{x}_{2N}$' and `$\bm{x}_{2N+1} \rightarrow \bm{x}_{N+1}$') is represented by the brown arrow. Assume the EVs to be uniformly distributed in the state intervals, a Markov transition matrix $\bm{A}\in \mathbb{R}^{(3N+3)\times (3N+3)}$ can be derived to model the state transition process. $\bm{A}_{m,n}$ denotes the transition probability from state $n$ to state $m$ every time step. It can be calculated analytically by EVs' inherent parameters \cite{wang2019state}. %\color{red} eSSM? check the other places \color{black}
The eSSM \color{black} can be structured as:
\vspace{-0.1cm}
\begin{equation}
\vspace{-0.1cm}
    {\footnotesize
        \begin{cases}
        \bm{x}(k+1) = \bm{A}\bm{x}(k)+\bm{B}\bm{u}^{\prime}(k) +\bm{w}(k)\\
        \bm{y}(k) = \bm{Cx}(k) +\bm{v}(k)
    \end{cases}
    }
    \label{eq:eSSM}
    \vspace{-0.1cm}
\end{equation}
In the eSSM, %\color{red} Did you really check???\color{black}, 
$\bm{u}^{\prime}(k)\in \mathbb{R}^{(4N+2)}\times 1$ is the input vector as
\vspace{-0.3em}
\begin{equation}
    {\footnotesize
        \bm{u}^{\prime}(k) = \begin{bmatrix}
        \bm{u}_{a}^\prime & \bm{u}_{b}^\prime& \bm{u}_{c}^{\prime} & \bm{u}_{d}^{\prime} & u_{d,N+1}^{\prime} & u_{b,N+1}^{\prime}
    \end{bmatrix}^{T}(k)
    }
\vspace{-0.2cm}
\end{equation}
where $\bm{u}_{a}^{\prime}$, $\bm{u}_{b}^{\prime}$, $\bm{u}_{c}^{\prime}$, and $\bm{u}_{d}^{\prime} \in \mathbb{R}^{1\times N}$ denote the control input vectors for %is the input vector for 
responding mode a - d in Fig. \ref{fig:states}. $u_{d,N+1}^{\prime}$ and $u_{b,N+1}^{\prime}$ %is the input element for the 
correspond to state transitions from $x_{3N+1}$ (IM at $S_{\min}$) to $\bm{x}_{1}$, % $u_{b,N+1}^{\prime}$ is the input element for the state transition 
and from $x_{3N+2}$ (IM at $S_{\max}$) %in IS ($\bm{x}_{3N+2}$) 
to $x_{3N}$, respectively. 
% The absolute value of each entry of the input vector represents the fraction of the total aggregated EV population that should switch. 
The absolute value of each input entry %entry's absolute value in the input vector 
represents the fraction of the total %aggregated 
EV population to switch between states %corresponding connecting states 
(see Fig. \ref{fig:state_transition}), and is bounded by the value of the corresponding state. %Hence, the range of each input element is constrained by the corresponding value of the states.

% \color{blue}Beside the input elements, control signals $\bm{u}$ that indicate the switch probability can be derived based on $\bm{u}^{\prime}$.
%\color{red} I don't understand this sentence. What is input elements, control signals. For (3), you call uprime to be input vector.\color{black}
%\color{blue}
$\bm{B}\in\mathbb{R}^{(3N+3)\times(4N+2)}$ is a constant matrix with each element
% (\ref{eq:B_value}). 
% \begin{equation}
% \vspace{-0.2em}
% \begin{small}
%             \bm{B} = 
% \begin{bmatrix}
%     -\bm{I}_{N\times N} & \bm{O}_{N\times N} & \bm{O}_{N\times N} & \bm{I}_{N\times N}  & \multirow{3}{*}{\makecell{1 \\ \vdots \\ 0}}& \multirow{3}{*}{\makecell{0 \\ \vdots \\ 1}}\\
%     \bm{I}_{N\times N} & -\bm{I}_{N\times N} & \bm{I}_{N\times N} & -\bm{I}_{N\times N} &  & \\
%     \bm{O}_{N\times N} & \bm{I}_{N\times N} & -\bm{I}_{N\times N} & \bm{O}_{N\times N} &  & \\
%     \\
%      \bm{O}_{1\times N} & \bm{O}_{1\times N} & \bm{O}_{1\times N} &  \bm{O}_{1 \times N} & -1 & 0 \\
%     \bm{O}_{1 \times N} & \bm{O}_{1\times N} & \bm{O}_{1 \times N} & \bm{O}_{1 \times N} & 0 & -1\\
%     \bm{O}_{1\times N} & \bm{O}_{1\times N} & \bm{O}_{1\times N} & \bm{O}_{1\times N} & 0 & 0
% \end{bmatrix}
% \label{eq:B_value}
% \end{small}
% % \vspace{-0.2em}
% \end{equation}
% %$\bm{B}\in\mathbb{R}^{(3N+3)\times(4N+2)}$ is a constant matrix as in (\ref{eq:B_value}). 
% $\bm{I}_{N\times N}$ is an $N\times N$ identity matrix; $\bm{O}_{m \times n}$ is an $m \times n$ zero matrix. 
$B_{m,n}$ indicates the influence of the control command $\bm{u}^{\prime}$ on the %\color{blue}
state variable %\color{red} state interval variable or state variable? \color{black} 
$x_{m}$, i.e. $x_{m}=\sum_{i=1}^{3N+3}A_{m,i}x_{i}(k)+\sum_{n=1}^{n=4N+2}B_{m,n}u_{n}^{\prime}$. $\bm{C} \in \mathbb{R}^{3\times (3N+3)}$ is a constant matrix determined by the number of connected EVs $N_{EV}$ and average rated charging/discharing power $P_{ac}$/$P_{ad}$ of EVs in CM/DM. Readers can refer to \cite{liu2025extended} for detailed derivations.
% (\ref{eq:C_value}).
% \begin{equation}
% \begin{small}
% \bm{C} = N_{EV}
% % \begin{align}
% \begin{bmatrix}
%     -P_{ac}\bm{1}_{N \times 1} & P_{ad}\bm{1}_{N \times 1} & -P_{ac}\bm{1}_{N \times 1} \\
%     \bm{O}_{N \times 1} & P_{ad}\bm{1}_{N \times 1} & -P_{ac}\bm{1}_{N \times 1} \\
%     P_{ad}\bm{1}_{N \times 1} & P_{ad}\bm{1}_{N \times 1} & -P_{ac}\bm{1}_{N \times 1} \\
%     0 & 0 & -P_{ac} \\
%     0 & P_{ad} & 0 \\
%     -P_{ac} &  -P_{ac} & -P_{ac}
% \end{bmatrix}^{T}
% % \end{align}
% \label{eq:C_value}
% \end{small}
% \end{equation}
% where $\bm{1}_{m\times n}$ is an $m\times n$ matrix of ones. $N_{EV}$ is the real-time total number of connected EVs. $P_{ac}$/$P_{ad}$ is the average rated charging/discharging power of EVs in CS/DS.
\color{black}

$\bm{y}(k)$ is the output vector as (\ref{eq:y_value}):
\vspace{-0.1cm}
\begin{equation}
\vspace{-0.2cm}
    {\footnotesize
        \bm{y}(k) = \begin{bmatrix}
        P(k) & \overline{P}(k) & \underline{P}(k)
    \end{bmatrix}^{T}
    }
    \label{eq:y_value}
    % \vspace{-0.1cm}
\end{equation}
where $P(k)$ is the power output of aggregated EVs. $\overline{P}(k)$/$\underline{P}(k)$ is the generation/absorption flexibility of the aggregated EVs, respectively. %\color{red} how is flexibility determined? \color{black}

$\bm{w}(k)$ is the combined process noise and modeling error, which also includes the random traveling behaviors (e.g., arrival, departure) of EVs\cite{wang2020electric}. It is assumed to be an independent Gaussian vector: $\bm{w}(k)\sim \mathcal{N}(\bm{0}, \bm{\Sigma}_{\bm{w}})$. $\bm{v}(k)$ is the combined measurement noise and modeling error, modeled as an independent Gaussian random vector: $\bm{v}(k)\sim \mathcal{N}(\bm{0}, \bm{\Sigma}_{\bm{v}})$. %\color{blue} The Gaussian assumption is applied only at the aggregate fleet level, not to individual EV dynamics. When the EV population is large and individual behaviors are weakly correlated, the aggregate residual can be viewed as the sum of many random effects. By the central limit theorem, its distribution approaches Gaussian even if individual components are discrete or non-Gaussian. In Section \ref{section:numerical validation}, we further test the robustness of the proposed method through simulations with non-Gaussian individual dynamics. \color{black}

%The innovations $\bm{w}(k)$ and $\bm{v}(k)$ capture the aggregated effect of many heterogeneous and weakly correlated uncertainties (e.g., plug-in/out events). For large $N_{EV}$ without strong synchronization, aggregation motivates an approximately Gaussian innovation model (by central-limit type reasoning), even if individual behaviors are non-Gaussian.
\color{black}

%\color{red} We haven't talked about parameter estimation yet here. I am not sure the computational burden is mainly about model matrix construction? prediction？or something else? \color{green} it's mainly about parameter prediction
%\color{blue}{
%It is worth noting that in eSSM, the computational burden of modeling and parameter estimation is determined primarily by the number of SOC bins $N$, rather than the number of connected EVs $N_{EV}$. Hence, for any fixed $N$, the computational burden remains essentially constant as the EV population grows. }

%The Gaussian assumptions are made only at the fleet (aggregate) level, not for individual EV dynamics. The innovations $\bm{w}(k)$ and $\bm{v}(k)$ capture the aggregated effect of many heterogeneous and weakly correlated uncertainties (e.g., plug-in/out events). For large $N_{EV}$ without strong synchronization, aggregation motivates an approximately Gaussian innovation model (by central-limit type reasoning), even if individual behaviors are non-Gaussian.
\color{black}
\color{black}

%\color{blue}
%\begin{equation}
%\bm{u}(k) = 
%\begin{bmatrix}
%    \bm{u}^{a}(k) & \bm{u}^{b}(k) & \bm{u}^{c}(k) & \bm{u}^{d}(k) & u^{d}_{N_{s}}(k) & u^{b}_{N_{s}}(k) 
%\end{bmatrix}^{T}
% \begin{bmatrix}
%     u^{a}(t)\\
%     u^{b}(t)
% \end{bmatrix}
%\label{eq:def_u}
%\end{equation}
%\color{black}

% \color{red} check the previous sentence, there are both "for instance" and "i.e." \color{black}

% \vspace{-1cm}
% \begin{figure}[t]
% \centering
% \includegraphics[width=1\columnwidth]{figs/ControlSignal.pdf}
% % \vspace{-0.2cm}
% %\caption{Solving PCCT by PCE method}
% % \vspace{-1cm}
% \caption{The State transition of aggregated EVs.} %\color{red} can you add the three additional states in the picture \color{black}}
% \label{fig:ControlS}
% % \vspace{-0.6cm}
% \end{figure}

\section{A novel %Generalizing 
Data-Driven Privacy-Preserving Model for Aggregated EVs} \label{section: bHMM}
%To capture the dynamics of EVs $\bm{x}(t)$, 
%\color{red} should be moved to the next section \color{black} 
The eSSM model %\color{red} the eSSM model doesn't appear before. Also, in the previous section, you need to make it clear what is SSM (previous proposed) and what is eSSM (you extended)  \color{black} 
can estimate the power outputs and flexibility of aggregated EVs efficiently. Nevertheless, it requires the knowledge of the real-time SOC values, traveling information (initial SOC, arrival/departure time) and characteristic parameters (e.g., the charging/discharging efficiencies, rated charging/discharging power, battery capacity) of all individual EVs. %\color{red} for the initial value of x?\color{black}, 
%as well as all the characteristic and traveling parameters (e.g., the charging and discharging efficiencies, battery capacity) of EVs. \color{red} tBD\color{black} 
This information is necessary to formulate the state variables $\bm{x}$ and parameter matrices $\bm{A}$ and $\bm{C}$. However, in reality, the collection of the above information is complex, privacy-intrusive, and most likely infeasible.

As a result, in this Section, we aim to develop a data-driven model that can %\color{blue} 
predict %\color{blue}
the flexibility and power outputs %\color{red}regulation capacity and flexibility? \color{black} 
of aggregated EVs %system output (total power output) $\bm{y}(t)$ %\color{red} can we say we capture the dynamics of x? \color{black} 
%from only the total power output $y(t)$ \color{blue} and other aggregated information 
without knowing SOC values, characteristics, and traveling parameters of individual EVs. 
This means predicting $\bm{Y}^{\prime}=\{\bm{y}(k+1), \cdots, \bm{y}(k+n_p\}$ ($n_p$ %\color{red} or $T_P$?\color{black}
is the prediction horizon) \color{black}from historical control input vectors along with past trajectories of power outputs from aggregated EVs. For example, the $l$-th trajectory takes the form 
\eqref{eq:dataset}: %\color{red} $\phi_L$ should be in bold \color{black}%\color{blue}
\vspace{-0.1cm}
\begin{equation}
\vspace{-0.3em}
{\footnotesize
    \bm{\phi}^{l}(k) = 
\begin{cases}
    \bm{U}^l \triangleq \{\bm{u}(k-K), \bm{u}(k-K+1), \cdots, \bm{u}(k-1)\}, \\
   \bm{Y}^l \triangleq \{P(k-K), P(k-K+1), \cdots, P(k)\}
\end{cases}
}
\label{eq:dataset}
\vspace{-0.4em}
\end{equation}
where $K$ is the length of the trajectory, such that%\color{red} will there be w in (11)?\color{black}
\vspace{-0.2cm}
\begin{equation}
\vspace{-0.1cm}
{\footnotesize
    \begin{cases}
\bm{\tilde{x}}(k+1) = \bm{\tilde{A}}\bm{\tilde{x}}(k) + \bm{\tilde{B}u}(k) + \tilde{\bm{w}}(k)\\ %\textcolor{blue}
%{+\bm{w}(k)}\\
\bm{y}(k) = \bm{\tilde{C}}\bm{\tilde{x}}(k)  + \bm{\tilde{v}}(k)
\end{cases}
}
\vspace{-0.3em}
\label{eq:systemidentification}
\end{equation}
%\color{red}
%This is a classical system identification problem. 
%The identified matrices.... and the true matrices... similar transformation (complete it)
%\color{blue} 
The identified matrices are a similarity transformation of the true matrices, which means they are input-output equivalent, with
%\color{black}To estimate the system from the dataset, the essential point is to determine a set of the hidden states $\widetilde{\bm{x}}$ and system parameters $[\bm{\widetilde{A}}, \bm{\widetilde{B}}, \bm{\widetilde{C}}]$ that are input-output equivalent to the original system $[\bm{A}, \bm{B}, \bm{C}]$ such that
\begin{equation}
\vspace{-0.3em}
{\footnotesize
        \begin{aligned}
        \bm{\tilde{A}} &= \bm{TAT}^{-1} \  \  \ \bm{\tilde{B}} = \bm{TB} \ \ \  \bm{\tilde{C}} = \bm{CT}^{-1} \\
        \bm{\tilde{x}} &= \bm{Tx} \ \ \ \ \ \ \  \ \ \tilde{\bm{w}} = \bm{Tw} \ \ \ \ \tilde{\bm{v}} = \bm{v}
    \end{aligned} 
    \label{eq:similarity transformation}
}
\vspace{-0.15cm}
\end{equation}
where $T\in \mathbb{R}^{(3N+3)\times (3N+3)}$ is an arbitrary invertible matrix. %\color{blue}Under this structure, the parameters we estimate from the inputs are a set of parameters after similarity transformation.

%develops a control framework that significantly reduces computational complexity by distributing EVs into different state intervals according to their connecting states and SOC values. \color{red} be more specific, \color{black} However, 
%the requirement for detailed information from EVs limits its practicability.
%This section describes the limitations of the SSM and develops a data-driven bilinear hidden Markov model (bHMM) for aggregated EVs. Besides, the EM algorithm is introduced to estimate the system parameters of the bHMM model.
% \vspace{-0.4cm}
\subsection{Formulation of a Bilinear Hidden Markov Model}

The eSSM %\color{red} change to eSSM thereafter \color{black} 
can be used for frequency regulation \cite{wang2019state} by designing appropriate $\bm{u}^{\prime}$ in (\ref{eq:eSSM}).
The design of $\bm{u}^{\prime}$ either requires direct access to or an accurate estimation of the 
state variables $\bm{x}$, as the range of input vectors is bounded by exact state values (i.e., $\bm{u}^{\prime}\in[0,\bm{x}]$). However,  %the system identification approach cannot give an exact estimation of 
accessing $\bm{x}$ may be privacy-intrusive, %the state variables $\bm{x}$ may not be accessed 
%restricted due to privacy constraints, 
and accurate estimation may be challenging %accurately estimated 
without feedback control\cite{verdult2002maximum} ($\bm{\widetilde{x}}=\bm{Tx} $ in (\ref{eq:systemidentification}) with $\bm{T}$ unknown). %\color{red} ref\color{black},
To eliminate the dependence of the control input $\bm{u}^{\prime}$ on $\bm{x}$, %address this issue, 
we replace %the input vector 
$\bm{u}^{\prime}$ by %some feasible control signal 
$\bm{u}$.
\vspace{-0.2cm}
\begin{equation}
\begin{small}
    \bm{u}(k) = 
\begin{bmatrix}
    \bm{u}_{a} & \bm{u}_{b} & \bm{u}_{c} & \bm{u}_{d} & u_{d,N+1} & u_{b,N+1}
\end{bmatrix} ^{T} (k)
\end{small}
\vspace{-0.3cm}
\label{eq:def_up}
\end{equation}
%\color{red} explain the detailed expression of $\bm{u}_p$. \color{black}  
where $u_{j}=u_{j}^{\prime}/x_{i} \ i=1,\cdots,3N+3$ %\color{red} why 4N+2? can I use $i$ rather than $j'$?\color{blue} 
and $i$ is the corresponding state interval subscript as shown in Fig. \ref{fig:state_transition}. As such,  $u_{j}$ represents %The value of each entry of the control signal represents 
the percentage of the EVs in $i^{th}$ state that should switch.
% the fraction of the corresponding interval population that should switch.
%\color{red}the percentage of the EV in $i$th state that should switch? \color{black}
%$\bm{u}_{p,j}=\bm{u}_{j}/\bm{x}_{j^{\prime}} \ j=1,\cdots,4N+2$ and $j^{\prime}$ is the corresponding state interval subscript as shown in Fig. \ref{fig:ControlS}. And 
Under this reformulation,
%that is not limited by the range of state values. Particularly, we let 
$\bm{u}$ is in the range of $[0,1]$, independent of $\bm{x}$. 

% A bilinear system such as \eqref{eq:bHMM} is a nonlinear extension of a linear system: the state evolution not only depends on the input and state but also their product. It is more general than linear systems but less complex than nonlinear systems. 

The extended SSM model (\ref{eq:eSSM}) thereby %new model 
can be reformulated as 
a %\color{blue} 
bilinear hidden Markov model (bHMM) %\color{red} Check i if BLS means bilinear system or bilinear state. I think it is inconsistent\color{black} 
\eqref{eq:bHMM}:
\vspace{-0.2cm}
\begin{equation}
\vspace{-0.4em}
{\footnotesize
%     \begin{cases}
% \bm{x}(k+1) = \bm{Ax}(k) + \sum\limits_{j=1}^{\rm{dim}\, \bm{u}}u_{j}(k)\bm{V}_{j}^{T}\bm{x}(k) + \bm{w}(k)\\
% \bm{y}(k) = \bm{C}\bm{x}(k) + \bm{v}(k)
% \end{cases}
\begin{cases}
\bm{x}(k+1) = \bm{Ax}(k) + \sum\limits_{j=1}^{{N_{\bm{u}}} }u_{j}(k)\bm{V}_{j}^{T}\bm{x}(k) + \bm{w}(k)\\
\bm{y}(k) = \bm{C}\bm{x}(k) + \bm{v}(k)
\end{cases}
}
\label{eq:bHMM}
% \vspace{-0.3em}
\end{equation}
where $\bm{A}$, $\bm{C}$ remain the same as in \eqref{eq:eSSM}; $N_{\bm{u}}$ is the dimension of input vector $\bm{u}$; %\color{red} eqn number, check all these details throughout the paper \color{black}
$\bm{V}=\{\bm{V}_{1},\cdots,\bm{V}_{N_{\bm{u}}}\} \color{black}$ is reformulated from $\bm{B}$ according to the new control vector $\bm{u}$ and takes the form in \eqref{eq:def_V} for $j =1, \cdots, N$: 
\vspace{-0.1cm}
%\color{blue} $\bm{u}_{p}$ is in the control signal sent to the EVs as \eqref{eq:def_up}. The value of each entry of the control signal represents the fraction of the corresponding interval population that should switch. $\bm{u}_{p,j}=\bm{u}_{j}/\bm{x}_{j^{\prime}} \ j=1,\cdots,4N+2$ and $j^{\prime}$ is the corresponding state interval subscript as shown in Fig. \ref{fig:ControlS}. And $\bm{V}$ is redistributed according to the new control vector as in \eqref{eq:def_V}.(explain the relation between up and u, as well as V). 
%\color{red} not clear what it means and how it is related to the development of the bhMM model. 
% \color{blue}
% In \cite{wang2019state}, after deriving the input vectors $\bm{u}$, the aggregator should drive a control signal that indicates the percentage of EVs in each state interval that should switch according to $\bm{u}$ and $\bm{x}$. The variation range of the control signal is $[0, 1]$. Replacing the input vector with the control signal, we can escape from the exact evaluation of state values and make system identification feasible.(add in the figure)\color{black}
% The nonuniqueness optimal $\theta$ of the full parameterization 
% By replacing the input vector $\bm{u}$ with control signal $\bm{u}_{p}$ as \eqref{eq:def_up}, we can derive a bHMM model. 
%\begin{equation}
%\bm{u}_{p}(k) = 
%\begin{bmatrix}
%    \bm{u}^{a}_{s} & \bm{u}^{b}_{s} & \bm{u}^{c}_{s} & \bm{u}^{d}_{s} & u^{d}_{s,N_{s}} & u^{b}_{s,N_{s}}
%\end{bmatrix} ^{T} (k)
%\label{eq:def_up}
%\end{equation}
\begin{equation}
% \vspace{-0.1cm}
\footnotesize
V_j =
\begin{cases}
[V_j]_{j,j} = -1,\; [V_j]_{j+N, j} = 1, & \text{CM} \!\to\! \text{IM},\\
[V_{j+N}]_{j+N, j+N} = -1,\; [V_{j+N}]_{j+2N, j+N} = 1, & \text{IM} \!\to\! \text{DM},\\
[V_{j+2N}]_{j+2N, j+2N} = -1,\; [V_{j+2N}]_{j+N, j+2N} = 1, & \text{DM} \!\to\! \text{IM},\\
[V_{j+3N}]_{j+N, j+N} = -1,\; [V_{j+3N}]_{j, j+N} = 1, & \text{IM} \!\to\! \text{CM},\\
[V_{4N+1}]_{3N+1, 3N+1} = -1,\; [V_{4N+1}]_{1, 3N+1} = 1, &  x_{3N+1} \\&\!\to\! x_1,\\
[V_{4N+2}]_{3N+2, 3N+2} = -1,\; [V_{4N+2}]_{3N, 3N+2} = 1, & x_{3N+2} \\&\!\to\! x_{3N}.
\end{cases}
\label{eq:def_V}
\end{equation}
% \begin{equation}
%     \begin{small}
%         \begin{aligned}
%         \begin{cases}
%         \begin{aligned}
%         \bm{V}_{j}\left[j,\, j\right] &= -1\\
%         \bm{V}_{j}\left[j+N,\, j\right] &= 1
%     \end{aligned}
%     &\quad{\text{CS} \rightarrow \text{IS}}\\
%     \\[2pt]
%     \begin{aligned}
%             \bm{V}_{j+N}\left[j+N,\, j+N\right] &= -1\\
%         \bm{V}_{j+N}[j+2N,\, j+N] &= 1
%         \end{aligned}
%         &\quad{\text{IS} \rightarrow \text{DS}}\\
%         \\
%     \begin{aligned}
%         \bm{V}_{j+2N}\left[j+2N,\, j+2N\right] & = -1\\
%         \bm{V}_{j+N}\left[j+N,\, j+2N\right] &= 1 \\
%     \end{aligned}
%     &\quad{\text{DS} \rightarrow \text{IS}}\\
%     \\
%     \begin{aligned}
%         \bm{V}_{j+2N}\left[j+N,\, j+N\right] &= -1 \\
%         \bm{V}_{j+2N}\left[j,\, j+N\right] &=1
%     \end{aligned}
%     &\quad{\text{IS} \rightarrow \text{CS}}
%     \end{cases}
%     \end{aligned}
%     \end{small}
%      \label{eq:def_V}
% \end{equation}
% Besides, $\bm{w}(t) \sim \mathcal{N}(0, \bm{\Sigma}_{w})$ is the measurement noise, and $\bm{v}(t) \sim \mathcal{N}(0, \bm{\Sigma}_{v})$ is the modeling error 
%\color{red} there is no modeling error in (4) (11). To be consistent, should we add the modeling error term in (4)?\color{black} %\color{blue}
Meanwhile, the initial conditions of the state variables (e.g., the percentage of EVs in each state interval) %historical trajectories \color{red} I think x are states not historical trajectorires\color{black} of aggregated EVs 
are assumed to %be subject a prior 
follow Gaussian distribution, i.e. $\bm{x}_{0} \sim \mathcal{N}(\bm{\mu}_{0}, \bm{\Sigma}_{0})$. As in the eSSM (\ref{eq:eSSM}),  $\bm{w}(k)\sim \mathcal{N}(\bm{0}, \bm{\Sigma}_{\bm{w}})$, $\bm{v}(k)\sim \mathcal{N}(\bm{0}, \bm{\Sigma}_{\bm{v}})$. These Gaussian assumptions are applied only at the aggregate fleet level, not to individual EV dynamics. When the EV population is large and individual behaviors are weakly correlated, the aggregate residual can be viewed as the sum of many random effects. By the central limit theorem, its distribution approaches Gaussian even if individual components are discrete or non-Gaussian. In Section \ref{section:numerical validation}, we further test the robustness of the proposed method through simulations with non-Gaussian individual dynamics. \color{black}

%\color{blue}Although the bHMM in~\eqref{eq:bHMM} includes state variables, these states represent aggregated population-level SOC intervals rather than individual EV states. Moreover, the model is identified solely from aggregated power measurements and control inputs, and the resulting parameters are input–output equivalent up to a similarity transformation, ensuring that individual EV SOC trajectories cannot be recovered.
%\color{black}

%\color{red} I think instead of discussing bilinear system, should change to bHMM \color{blue}
Bilinear hidden markov models (bHMMs) represent one of the simplest classes of nonlinear systems and are therefore often used as a tractable framework for analyzing more complex nonlinear dynamics \cite{pardalos2010optimization}. %\color{blue} 
In a bHMM, \color{black}the state evolution is linear in the state and input individually but nonlinear in their joint interaction through state–input products. This structure naturally models scenarios where control actions multiply with system states, making it well suited for data-driven EV control.\color{black}
% \color{black}$\bm{u}^{a}_{s}$, $\bm{u}^{b}_{s}$, $\bm{u}^{c}_{s}$, and $\bm{u}^{d}_{s}\in \mathbb{R}^{N\times 1}$ are new input vectors. $u_{s, N_{s}}^{b}$ and $u_{s, N_{s}}^{d}$ are special extended input bins designed for $\bm{x}_{3N_{s}+1}$ and $\bm{x}_{3N_{s}+2}$. 
% Their respective responding modes are shown in Fig. \ref{fig:ControlS}.

% The bHMM can be written as \eqref{eq:bHMM}:
% \begin{equation}
% \begin{cases}
% \bm{x}(k+1) = \bm{Ax}(k) + \sum\limits_{j=0}^{\rm{dim}\, \bm{u}}\bm{u}_{p,j}(k)V_{j}^{T}\bm{x}_{k}(k) + \bm{w}(k)\\
% y(k) = \bm{C}\bm{x}(k) + v(k)
% \end{cases}
% \label{eq:bHMM}
% \end{equation}
% where $\bm{A}$ is the transition matrix. 

% For that purpose, we apply bHMM model...... 
% Specifically, instead of using $\bm{u}(t)$ that is the proportion of EVs among all EVs that should be switched between different states, we define $\bm{u}_{p}(t)$ as in \eqref{eq:def_up}, which describes the probability of EVs that should switch between different states. Consequently, the variation range of $\bm{u}_{p}(t)$ %changes to becomes $[0, 100\%]$. 

%\color{blue} 
By taking the observed trajectories and inputs as if they come from a bHMM whose hidden dynamics are unkown, we avoid specifying detailed SOC distributions to form $\bm{x}$. %\color{blue} thereby preventing leakage
%of private user information. \color{black}
The goal now is to estimate the model parameters 
\vspace{-0.2cm}%problem now turns into estimating the model parameters %\color{red} give all the parameters\color{black} \color{blue} and initial conditions \color{red} give the detailed notation\color{black}%describing the model
%\color{blue} 
%s shown  in \eqref{eq:parameters}:
%\color{blue}
\begin{equation}
\vspace{-0.2cm}
{\footnotesize
    \bm{\Theta} = \{ \bm{A},\bm{V}, {\bm{C}_1}, \bm{\Sigma}_{\bm{w}}, \bm{\Sigma}_{\bm{v}}, \bm{\mu}_{0}, \bm{\Sigma}_{0} \}
    }
    \label{eq:parameters}
\end{equation}\color{black}
from historical \color{black}observations of input $\{\bm{u}_k \}$ and output $\{P_k\}$ in \eqref{eq:dataset}. $\bm{C}_1$ is the first row of matrices $\bm{C}$. %with no access to the SOC states of EVs. %To estimate the parameters of the bHMM, 
To this end, the next section introduces an %\color{red} give the full name when it first appear. Check if EM first appear \color{blue} it first appears in the introduction\color{black} 
EM-based algorithm %will be proposed in the next section, which can 
that predicts the %\color{blue} 
total power outputs and flexibility of %\color{red} can it predict x? \color{black} 
of EVs from only the aggregated EV data and control inputs %of aggregated EVs 
without requiring individual EV data. To ensure accuracy, the model parameters are updated periodically using a moving window.

\color{black}  %power trajectories of historical data. 
%is implemented to estimate the parameters \{A, C\} to predict the state space of EVs from the power trajectories of historical data.
% Compared with the SSM, the bHMM also simplifies the calculation of regulation capacity. 
% The upper boundary can be estimated by setting all elements of $\bm{u}_{p}^{CI}(t)$ and $\bm{u}_{p}^{ID}(t)$ to 1; similarly, setting $\bm{u}_{p}^{DI}(t)$ and $\bm{u}_{p}^{IC}(t)$ to 1, the lower boundary can be estimated. The calculation of boundaries in the SSM requires the whole information of states and parameters.
%\color{red} not clearly explained. will discuss\color{black}
% \color{red} how does (4) calculate the regulation capacity?\color{black}
% \vspace{-0.2cm}

\subsection{The Expectation-Maximization Algorithm}
% \vspace{-0.1cm}
% This section provides a preliminary introduction to the key techniques in this paper, i.e. bilinear model and EM algorithm.
%\color{red} still not very clear why you use bilinear model? Will this kind of bilinearity help capture more complex/nonlinear behavior? why is it needed? 
%A multivariable bilinear system can be identified by optimizing the output-error cost function. 
Although the eSSM model in the form of (\ref{eq:bHMM}) eliminates the dependence of $\bm{u}$ on $\bm{x}$, it is expressed as a bHMM, unlike the original linear model in (\ref{eq:eSSM}), making it challenging to identify model parameters. Existing bHMM identification methods, such as gradient-based \cite{li2017gradient} and subspace approaches \cite{favoreel1999subspace}, often lack scalability and are computationally intensive for large systems. In contrast, maximum likelihood methods, while slower to converge, offer better numerical stability, reliable convergence, and simpler implementation.
%\color{blue}
% Some identification methods for the bHMM %bilinear systems 
% have been proposed in the literature, ranging from gradient-based methods%\color{blue}
% \cite{li2017gradient} %\color{red} cite\color{black} 
% to subspace methods %\color{blue} 
% \cite{favoreel1999subspace}. %\color{red} cite\color{black}.
% %Gradient-based search methods show high efficiency on estimates of state space problems.
% %\color{red} system identification?\color{blue} 
% However, they do not have good scalability and suffer from high computational demands as the system size increases. %\color{red} cite if possible\color{black}
% In contrast, maximum likelihood methods, though slower in terms of convergence rate, offer greater numerical stability, reliable convergence, and simplicity of implementation\cite{ng2012algorithm}. %\color{red}

Hence, this paper explores the application of EM algorithm—a widely used maximum likelihood approach for hidden state estimation. %...\color{blue} intends to 
%explores the application of maximum likelihood methods to EV data. %highlighting its advantages in handling incomplete datasets often encountered in this field. \color{black} %\color{red} please cite a ref to support this claim. \color{black}%there are too many terminologies, do you mean EM is low in convergence? what do you mean by reliability? how is it defined? \cite{gibson2005maximum}
 % \color{red} how is missing data related to the problem you study here? 
%\color{blue}
%The EM %\color{red} give its full name when it first appear in the paper \color{black}
%algorithm is a broadly applicable maximum likelihood method used for %\color{blue} 
%hidden Markov state estimation problems. %\color{red} The incomplete-data is a confusing and vague expressing. What data are you referring to?\color{black}
By taking this approach, the values of hidden states (i.e., $\bm{x}$ in (\ref{eq:bHMM})) can be estimated during the expectation step (E-step) with an optimal state estimator. The maximization step (M-step) updates model parameters using these estimates.
% It unravels the complex dependence between the observed output sequences, initial states, and time-varying input sequences.\color{red} what does the previous sentence mean?\color{blue} 
%The maximization (M) step then utilizes the %\color{red} what is untangled? \color{blue}
%estimated states to %compute 
%update parameter approximations.
% Its basic idea is treating the observed dataset as if it comes from a hidden Markov model and estimating parameters after filling in initial values for the missing data. The initial values can then be updated by the predicted parameters. Then the parameters are re-estimated, proceeding iteratively until convergence. This process is composed of two repetitive steps: the expectation step (E-step) and the Maximization step (M-step). 

% which means the state distribution and characteristic parameters we need to construct $x(t)$ and model parameters.
%\color{red} check the previous sentence. not clear \color{black} 
% It treats the observed trajectories as if t and estimates parameters by iteratively computing the Expectation step (E-step) and the Maximization Step (M-step).
% It decomposes the problem into two steps: the Expectation Step (E-step) and the Maximization Step (M-step)\cite{ng2012algorithm}.

%\color{blue}
In our case, both EV parameters and hidden states $\bm{x}$ are unknown. %For our missing 
%the EV parameters and states $\bm{x}$ are unknown, %case \color{red} what data is missing? \color{black}, 
%the solving process with EM algorithm is as follow. %\color{red} merge with the previous paragraph \color{black} 
%Assuming %\color{blue} 
At each time step \color{black} we assume access to a dataset $\bm{\Phi}(k)$ (see (\ref{eq:dataset})) as in \eqref{eq:dataset_full}: 
\vspace{-0.1cm}%\color{red} notation inconsistent from (7) \color{black}
\begin{equation}
\vspace{-0.2cm}
    {\footnotesize
        \bm{\Phi}(k) = \{\bm{\phi}^{1}(k), \bm{\phi}^{2}(k), \cdots, \bm{\phi}^{L}(k)\}
    }
    \label{eq:dataset_full}
    % \vspace{-0.1cm}
\end{equation}
which consists of $L$ independent trajectories of aggregated EV power data (including flexibility, power outputs, and control inputs) from the same time window as in~\eqref{eq:dataset}, each of length $K$. The dataset can be constructed using data from $L-1$ previous days and the current prediction day. \color{black}%\color{red} each with length k? \color{black} %\color{red} why do you use K in (10), but L here? Also should it be $\bm{\Phi}$ in (10)?\color{black} %\color{red} why do you use $\phi$? not the original $P_{EV}$?\color{black}
% ; the prior Gaussian distribution for the initial condition is $\bm{x}_{0} \sim \mathcal{N}(\bm{\mu}_{0}, \bm{\Sigma}_{0})$. 
%\color{blue}
The goal is to estimate model parameters $\bm{\Theta}$ in \eqref{eq:parameters} by maximizing the log-likelihood of the observed data 
$\bm{\Phi}(k)$ \eqref{eq:dataset_full}, leveraging the Markov structure of the model in \eqref{eq:bHMM}, as shown in \eqref{eq:log_likelihood}.
%The maximum likelihood estimation aims to maximize the probability of the observed data \eqref{eq:dataset_full} %\color{red} or (15)?\color{black} 
%over the model parameters \eqref{eq:parameters}. Because of the Markov property of the model, \eqref{eq:bHMM}, %\color{red} is it Markov?\color{black}, 
%we aim to maximize % which can be represented by a 
%the log-likelihood function based on the collected dataset $\bm{\Phi}$ as shown in \eqref{eq:log_likelihood}: %\color{red} Is there a particular reason for taking the log-likelihood?\color{black}
\vspace{-0.2cm}
\begin{equation}
\vspace{-0.1em}
{\footnotesize
    L(\bm{\Theta}) = \sum _{l=1}^{L} \log P_{\bm{Y}}(\bm{Y}^{l}; \bm{\Theta})
}
\label{eq:log_likelihood}
\vspace{-0.2cm}
\end{equation}
% The prior Gaussian distribution for the initial condition is . Let $\bm{V}=\{\bm{V}_{j}, j=1,\cdots,4N+2\}$, the parameters $\Theta$ of the model \eqref{eq:bHMM} that we seek to determine are defined as \eqref{eq:parameters}:
% The log-likelihood of the observations can be denoted by \eqref{eq:log_likelihood}:
% where $\bm{Y}^{(l)} = \{ y_{k}^{(l)} \}_{0\leqslant k \leqslant K}$, 
where $P_{\bm{Y}}(\bm{Y}^{l}; \bm{\Theta})$ denotes the probability density of observing trajectory $\bm{Y}^{l}$ given model parameters $\bm{\Theta}$, computed as \eqref{eq:Density_Y}:
\vspace{-0.1cm}
\begin{equation}
{\footnotesize
\begin{aligned}
P_{\bm{Y}}(\bm{Y}^{l}; \bm{\Theta}) &= \int P_{\bm{X, Y}}\bm(X, Y^{l}; \bm{\Theta})d\bm{X}= \mathbb{E}_{\hat{\bm{X}}^{l}}\left( \frac{P_{\bm{X, Y}}( \bm{\hat{X}}^{l}, \bm{Y}^{l}; \bm{\Theta})}{Q( \bm{\hat{X}}^{l})}  \right)
\end{aligned}
% \vspace{-0.4em}
}
\label{eq:Density_Y}
\end{equation}
%\color{red} check the original paper and see how the rationale is explained \color{black} 
$P_{\bm{X, Y}}$ is the joint density of the hidden %\color{red} hidden or laten? \color{black} 
state variables $\bm{X}=\{ \bm{x}_{k} \} _{k=0}^{K},l=1,\cdots, L$ and the observed trajectory $\bm{Y}^l$ (see \eqref{eq:dataset}). %\color{blue} Y is defined in (7)\color{red} Is Y defined? \color{black} 
$\bm{\hat{X}}$ is an introduced random variable. $\bm{X} \longmapsto Q^{l}(\bm{X})$ is an introduced probability density, %which is referred to
referred as the ``inference" distribution of the %\color{blue} 
hidden %\color{black} 
states $\bm{x}$. %\color{red} hidden or latent? may use one consistently \color{black}
% \color{red} It seems $Z$ should be changed to $X$?\color{black}

% \color{red} needs to explain why you can use log. \color{black}
% The log function is a commonly used concave function\cite{bagnoli2006log}.
%\color{blue}
Due to the concavity of the log function, Jensen’s inequality can be applied to the log-likelihood in  \eqref{eq:log_likelihood}, yielding a lower bound commonly known as the Evidence Lower Bound (ELBO): %Since the log function is a %common 
%concave function, we can apply %\color{red} essential information that the log is a concave function is missing \color{black} 
%Jensen's inequality to the \eqref{eq:log_likelihood}, a lower boundary of the log-likelihood, which is usually referred to as Evidence Lower Bound (ELBO), can be constructed as \eqref{eq:ELBO}:
\begin{equation}
% \vspace{-0.3cm}
{\footnotesize
\begin{aligned}
L\left( \bm{\Theta} \right) 
&\geqslant \hat{L}_{Q}(\bm{\Theta}) \\
&= \sum_{l=1}^{L} \bigg\{ 
    \mathbb{E}_{\hat{\bm{X}}^{l}} \left[ 
        \log P_{\bm{X,Y}} \left( \bm{\hat{X}}^{l}, \bm{Y}^{l}; \bm{\Theta} \right) 
    \right] 
    - \mathbb{E}_{\hat{\bm{X}}^{l}} \left[ 
        \log Q^{l}(\bm{\hat{X}}^{l}) 
    \right] 
\bigg\} \\
&= L(\bm{\Theta}) - 
\sum_{l=1}^{L} \mathbb{E}_{\hat{\bm{X}}^{l}} \left[ 
    \log \left( 
        \frac{Q^{l}(\bm{\hat{X}}^{l})}
             {P_{\bm{X}|\bm{Y}=\bm{Y}^{l}}(\bm{\hat{X}}^{l}; \bm{\Theta})} 
    \right) 
\right]
\end{aligned}
}
\label{eq:ELBO}
\end{equation}
%\color{red} 
A key observation is \color{black} that the inequality takes equality, i.e. 
% \vspace{-0.2cm}
\begin{equation}
% \vspace{-0.1cm}
{\footnotesize
    L(\bm{\Theta}) = \hat{L}_{Q}(\bm{\Theta})
}
\label{eq:ELBO equality}
% \vspace{-0.3em}
\end{equation} 
when the probability density $Q^{l}$ is the conditional density of $\bm{X}$ given $\bm{Y}^{l}$, i.e. $Q^{l} = P_{\bm{X}|\bm{Y}=\bm{Y}^{l}}$. 
%Equation \eqref{eq:ELBO} and its conditions for taking equal signs 
This principle implies the EM algorithm's ascent optimization approach for maximizing the likelihood: Given an initialization parameter set $\bm{\Theta}_{0}$, the EM algorithm iteratively updates the inference distributions $Q^{l}$ with the parameters $\bm{\Theta}$ fixed and then updates the parameter $\bm{\Theta}$ with the inference distribution $Q^{l}$ fixed. %\color{blue} whether takes E step or M step first are both Ok \color{red} check, the order is different from maximization step and expectation step\color{black}. 
Given an initial parameter estimate $\bm{\Theta_0}$, this process produces updated parameters  $\bm{\Theta_1}$:
% \vspace{-0.2cm}
% Assuming $\bm{\Theta}_{1}$ to be the new model parameters found by the maximization step, we have: 
\begin{equation}
\vspace{-0.3em}
    {\
        L(\bm{\bm{\Theta}_{1}}) \geq \hat{L}_{Q}(\bm{\Theta}_{1}) \geq \hat{L}_{Q}(\bm{\Theta}_{0}) = L(\bm{\Theta}_{0})
    }
    \label{eq:likelihood}
    \vspace{-0.2em}
\end{equation}
\vspace{-0.3em}
where the first inequality is from Jensen's inequality \eqref{eq:ELBO}, the second is from the %inequality is determined by the 
M-step, and the equality is by the definition of $Q^{l}=P_{\bm{X}|\bm{Y}=\bm{Y}^{l}}$ and (\ref{eq:ELBO equality}). This process ensures that the log-likelihood increases with each iteration \cite{otto2022learning}. 
%In this process, the log-likelihood function \eqref{eq:log_likelihood} is guaranteed to increase at each step\cite{otto2022learning}.  
It can be decomposed into two steps:

(1) E-step: update $Q^{l}$ with the parameter $\bm{\Theta}$ fixed, and calculate $\hat{L}(\bm{\Theta})$. For each $l$, set
\vspace{-0.1cm}
\begin{equation}
\vspace{-0.2cm}
{\footnotesize
    Q^{l}(\bm{\hat{X}}^{l}) := P_{\bm{X,Y}}(\bm{Y}^{l}|\bm{\hat{X}}^{l};\bm{\Theta})
}
\label{eq:e_step}
% \vspace{-0.1cm}
\end{equation}
This is achieved by the Kalman filter and smoother. Readers can refer to details in Appendix A. %\color{black}

(2) M-step: update the parameter $\bm{\Theta}$ with fixed ``inference'' distributions $\{Q^{l}\}_{l=1}^{L}$. Set 
% \color{red} what is ELBO? \color{black}
\vspace{-0.25cm}
\begin{equation}
\vspace{-0.1cm}
    {\footnotesize
        \begin{aligned}
            \bm{\Theta}
            &= \mathop{\arg\max}\limits_{\bm{\Theta}} \sum_{l=1}^{L} \mathrm{ELBO}(\bm{\hat{Y}}^{l} \mid \bm{\hat{X}}^{l}; \bm{\Theta}) \\
            &= \mathop{\arg\max}\limits_{\bm{\Theta}} \sum_{l=1}^{L} \mathbb{E}_{\bm{\hat{X}}^l}\bigg[ \log \frac{P_{\bm{X}, \bm{Y}}(\bm{\hat{X}}^{l}, \bm{Y}^{l}; \bm{\Theta})}{Q^{l}(\bm{\hat{X}}^{l})} \bigg]
        \end{aligned}
    }
\label{eq:m_step}
\vspace{-0.1cm}
\end{equation}
According to \cite{otto2022learning}, we can derive an analytical solution for \eqref{eq:m_step} through \eqref{eq:optimal_mu} to \eqref{eq:cal_Hl} %\color{red} correct equation number \color{black}... 
in Appendix B. 

%\color{blue}
Repeating these two steps guarantees a monotonic increase in the log-likelihood at each iteration~\cite{otto2022learning}. The convergence of this process has been formally analyzed in~\cite{wu1983convergence}, which shows that the algorithm converges under mild regularity assumptions—namely, a well-defined and continuously differentiable likelihood function as~\eqref{eq:likelihood}, the existence and finiteness of the expectation as~\eqref{eq:ELBO} in the E-step, and local maximization of the auxiliary function~\eqref{eq:m_step} in the M-step (analytical solutions for~\eqref{eq:m_step} are derived in Appendix B)—all of which are satisfied in this work. %\color{red} assumptions conditions? \color{black}. 
But one limit of the EM algorithm is that we only find a local minimum. %\color{blue} Selection of proper initialization paramter set can enhance our prediction results. To improve the algorithm's ability to capture the system's features, we utilized the system structure and some aggregated information %\color{red} will there be more detailed description in the algorithm about the aggregated information? \color{black} 
%to initialize the parameter set $\bm{\Theta}^{(0)}$ as shown in remarks.

After the parameter set is identified, it can be used to predict the aggregate power output and the flexibility. Under the proposed formulation, flexibility bounds are obtained by appropriately selecting the control variables that activate admissible state transitions. To compute the upper flexibility bound $\overline{P}$, all entries of $\bm{u}_a$, $\bm{u}_b$, and $u_{b, N+1}$  are set to be $1$ in (\ref{eq:def_up}), which means activate all admissible transitions that switch EVs across different states in the directions as illustrated in Fig. 2. The lower flexibility bound $\underline{P}$ is obtained by setting every entry of $\bm{u}_c$, $\bm{u}_d$, and $u_{d, N+1}$ to $1$ in (\ref{eq:def_up}), which enforces the opposite transitions. These control settings are used only in simulations to estimate flexibility. %This construction provides a systematic way to estimate the aggregated power output  bounded aggregated power responses with feasible state evolutions. 

The detailed algorithm for predicting aggregated power outputs and flexibility is outlined in Algorithm \ref{alg:EM}. 
\color{black}
\begin{algorithm}[t]
\LinesNotNumbered
\small
\caption{Aggregated Power and Flexibility Estimation }
\SetAlgoNlRelativeSize{-0.5}
%\color{blue}$\bm{\Phi}(k), 
    %$\epsilon_{\min}$,
   % $N_{ITER}$)}%\color{red}I don't know your rules separating inputs and parameters, Are U, Y inputs or parameters? \color{blue} inputs are what we have, parameters are what we need to estimate.U and Y should be inputs here.\color{black}}
    \label{alg:EM}
    \KwIn{A dataset $\Phi$ with $L$ independent trajectories (including 
$L-1$ historical and real-time) of control inputs, aggregated power outputs, 
% and \color{red} the estimated flexibility ?\color{black}; %\color{blue} power trajectories and flexibility \color{red} do you need flexibility?\color{black}; 
initial parameter set $\bm{\Theta}^{(0)}$; convergence threshold $\epsilon_{min}$; maximum number of iterations $N_{ITER}$}
    \KwOut{bHMM parameter set $\bm{\Theta}$; predicted aggregated power outputs and generation/consumption flexibility  $\bm{y}(m)=[P(m), \overline{P}(m), \underline{P}(m)]^T$ for %the next %\color{blue}
   %$n_p$ \color{black}intervals, i.e., 
   $m=k+1,k+2,...,k+n_p$} %\color{red} check here as well \ color{black}}
    % s $\Theta = \{ \bm{A}, \bm{C}, \bm{\Sigma}_{\bm{w}}, \bm{\Sigma}_{\bm{v}}, \bm{\mu}_{0}, \bm{\Sigma}_{0} \}$}
    \textbf{Initialization}:\\ %Initialize the parameter 
    Set $\bm{\Theta}^{(0)} = \{ \bm{A}^{(0)}, \bm{C}_{1}^{(0)}, \bm{\Sigma}_{\bm{w}}^{(0)}, \bm{\Sigma}_{\bm{v}}^{(0)}, \bm{\mu}_{0}^{(0)}, \bm{\Sigma}_{0}^{(0)} \}$, %and
    $n=0$.
    
    % \color{red} what distributions. how? describe specifically \color{black}
    % , and determine the minimum likelihood iteration error $\epsilon_{min}=10^{-2}$ and maximum iteration number $N_{ITER}$
    \While{ $\epsilon > \epsilon_{min}$ \textbf{or} $n \leq N_{ITER}$}{
        E-step: Estimate posterior distribution $\bm{Q}^{(n)}$ using \eqref{eq:e_step}.\\
        M-step: Update parameters $\bm{\Theta}^{(n)}$ using \eqref{eq:m_step};\\
        Calculate log likelihood $L(\bm{\Theta}^{(n)})$ via \eqref{eq:log_likelihood};\\
        Set $\epsilon = L(\bm{\Theta}^{(n)}) - L(\bm{\Theta}^{(n-1)})$ and $n = n + 1$.\\
    } 
    Estimate the aggregated power and flexibility using $\Theta^{(m)}$ and \eqref{eq:def_up}- \eqref{eq:bHMM}. Particularly, set all entries of $\bm{u}_a$, $\bm{u}_b$ and $u_{b, N+1}$ as $1$ to obtain $\overline{P}(m)$; set all entries of $\bm{u}_c$, $\bm{u}_{d, N+1}$ as 1 to obtain $\underline{P}(m)$. %\color{red} explain clearly setting what signals to 1 to get upper bound and what signals to 1 to get lower bound. 
    
    %\color{blue} 
    \textbf{Return} Final model parameters %\color{red} check and modify \color{black}
    $\bm{\Theta}=\bm{\Theta}^{(n)}$. %, which is the final estimated model parameters. % for output. \color{black}
    The estimated $\bm{y}(m)=[P(m), \overline{P}(m), \underline{P}(m)]^T$% in  \eqref{eq:bHMM} 
    for $m=k+1,...k+n_p\color{black}$.
    % Use  $\bm{\Theta}$ to calculate $\bm{y}(m)=[P(m), \overline{P}(m), \underline{P}(m)]^T$ in  \eqref{eq:bHMM} for $m=k+1,...k+n_p\color{black}$. %in \eqref{eq:bHMM}, where $
        %\bm{y}(k)=[P(k), \overline{P}(k), \underline{P}(k)]^T$, $k=1,2,...n_p$, which contains the estimated aggregated  power $P(k)$ and the generation/consumption flexibility $[\overline{P}(k), \underline{P}(k)]$ for the next $n_{p}$ time intervals. %\color{red} Is $\Theta^{(m)}$ the final $\Theta$? \color{black}
    %\textbf{return} Predicted aggregated power and flexibility
    % Forecast load and flexibility with 
\end{algorithm}

\textbf{Remarks:}\\
$\bullet$ Proper initialization of model parameters can significantly improve prediction accuracy.  %Selection of proper initialization parameter set can enhance prediction results. %To improve the algorithm's ability to capture the system's features, 
We leverage the system structure and some aggregated information %\color{red} will there be more detailed description in the algorithm about the aggregated information? \color{black} 
to initialize the parameter set $\bm{\Theta}^{(0)}$. %as shown in remarks.
Specifically, $\bm{C}_{1}^{(0)}$ is initialized as \eqref{eq:C_initial}:  
\begin{equation}
% \vspace{-0.1cm}
\label{eq:C_initial}
{\footnotesize
\bm{C}_{1}^{(0)} = \eta
% \begin{align}
\begin{bmatrix}
    -\bm{1}_{1\times N} & \bm{O}_{1\times N} & \bm{1}_{1\times N} & 0 & 0 & -1\\
    % \bm{1}_{1\times N} & \bm{1}_{1\times N} & \bm{1}_{1\times N} & 0 & 1 & -1\\
    % -\bm{1}_{1\times N} & -\bm{1}_{1\times N} & -\bm{1}_{1\times N} & -1 & 0 & -1 \\
\end{bmatrix}
% \end{align}
}
\end{equation}
where $\eta\sim U(N_{EVs}P_{\min}, N_{EVs}P_{\max})$,  %is a random variable generated by a uniform distribution $U(N_{EV}P_{\min},  N_{EV}P_{\max})$ 
and $N_{EVs}$ is the total number of aggregated EVs. %\color{blue}
$P_{\min}/P_{\max}$ is the estimated minimum/maximum power of the aggregated EVs that can be deduced from historical aggregated data. %literature.\color{red} did you explain $P_{min}$ $P_{max}$?\color{black}%\color{red} $N_{EV}$ and $N_{EV}^{\prime}$?\color{black}%\color{red} 
%In theory, \color{blue} $\bm{\mu}_{0}^{(0)}$ can be initialized arbitrarily; in practice, ... %\color{blue}
%we find that the algorithm works better when we consider that 
$\bm{V}^{(0)}$ is initialized as \eqref{eq:def_V}. These choices encode the known physical structure of the aggregated fleet model before learning, aiming to improve the parameter estimation results.%and prevent the initialization from injecting an unintended prior that could bias the early EM iterations toward particular state-transition trajectories. \color{black} %In addition, the model structure itself incorporates physical constraints that restrict infeasible transitions
%(e.g., charging/discharging directionality, boundary-state restrictions), ensuring that even under uniform initialization, the parameter space remains physically admissible. Therefore, impossible state transitions (e.g.,
%skipping SOC intervals or violating boundary dynamics) are structurally prevented. \color{black} %Since 
The elements of $\bm{\mu}_{0}^{(0)}$, representing the percentage of EVs in the state bins, are drawn from $U(0,1)$ to ensure values in $[0,1]$.  % and should be in the range $[0, 1]$. The elements of $\bm{\mu}_{0}^{(0)}$ is randomly generated from the uniform distribution $U(0,1)$. 
Covariance matrices $\bm{\Sigma}_{0}^{(0)}$, $\bm{\Sigma}_{\bm{w}}^{(0)}$, and $\bm{\Sigma}_{\bm{v}}^{(0)}$ are initialized as $\beta \bm{I}_{M\times M}$, where $M$ is matrix dimension and %the dimension of the respective matrix and 
$\beta\sim U(0,1)$. \\%is generated from the uniform distribution $U(0,1)$.\\

To evaluate robustness, we repeated the estimation with 10 %\color{red} give a number \color{blue} 
random seeds using the same initialization strategy. %The resulting parameter estimates and reconstruction errors were nearly identical across runs, 
Results indicate that the proposed initialization consistently leads to stable convergence and accurate estimation performance, whereas random initialization may lead to lower accuracy. % while random initialization can lead to low accuracy. 
\color{black}%We have performed.... repeated run using the same initialization strategy,....simulation results indicate that this initialization strategy consistently leads to accurate estimation performance.
%The large-population nature of aggregated EV systems further mitigates the sensitivity of EM to initialization. When many EVs are aggregated, individual stochastic variations (e.g., arrival times and SOC levels) average out, leading to smoother aggregate power dynamics. This aggregation effect regularizes the likelihood surface of the EM optimization, making extreme local optima less likely. Consequently, different admissible initializations $\Theta^{(0)}$ typically yield similar convergence trajectories and comparable parameter estimates. 
% We verified this by repeating the identification with several randomly generated initial parameter sets $\Theta^{(0)}$, which resulted in nearly identical convergence behavior and reconstruction errors.
%\color{red} how did you verify this?\color{black}

\noindent$\bullet$ The value of $n_p$ affects the prediction accuracy.
In our simulations, with $\Delta t= 15s$, $n_p=12$, the same parameters are used over a 3 mins horizon for prediction, yielding reasonably accurate results consistent with the observations in ~\cite{wang2020electric}. 

\begin{figure}[t]
% \vspace{-0.3em}
    \centering
    \includegraphics[width=0.9\linewidth]{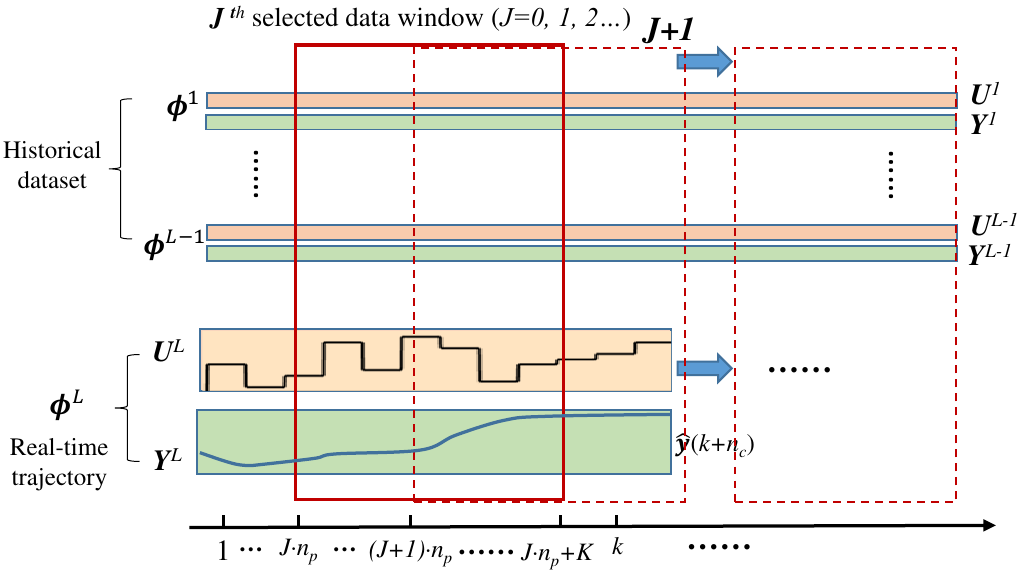}
    % \vspace{-0.3cm}
    \caption{Periodic updates of model parameters using sliding windows.}
    \label{fig:MPC}
    % \vspace{-0.4em}
\end{figure}

\noindent$\bullet$ 
%It should be noted that t
The estimation accuracy of Algorithm 1 may decrease over time due to, %\color{blue} 
EVs transitioning between states due to natural dynamics, control actions, and their traveling behaviors. To maintain accuracy, we adopt a sliding-window estimation strategy (see Fig.~\ref{fig:MPC}). A shorter window improves tracking of time-varying dynamics (e.g., EV arrivals and departures) but increases estimation noise, while a longer window enhances stability at the expense of slower adaptation. We use a 15-minute window (i.e., each  trajectory has length $K=60$ under $\Delta t=15s$) with 3-minute rolling updates to balance robustness and adaptability, meaning that parameters are updated every 3 minutes using the most recent 15 minutes of aggregated data.  %enabling adaptation to changing EV populations and operating conditions while maintaining estimation stability through sufficient averaging. 

\noindent $\bullet$ 
Identifiability of the bHMM parameters requires the empirical information matrix in the M-step update \eqref{eq:cal_V} to be nonsingular (see Appendix B), corresponding to the classical persistent excitation condition~\cite{ioannou1996robust}. This requires sufficient temporal excitation within trajectories and diversity across trajectories.
%This requirement depends on both temporal richness within each trajectory and diversity across trajectories. 
Increasing the trajectory length $K$ improves temporal richness, while increasing the number of trajectories $L$ improves ensemble diversity under varying operating conditions. Larger $L$ and $K$ also increase the effective sample size $L(K+1)$, thereby reducing estimation variance. However, excessively large $K$ may reduce adaptability under a sliding-window scheme, and large $L$ increases computational cost. %If there are sufficiently rich excitation, then the nonsingularity of \eqref{eq:cal_V}  is almost surely if $L(K+1)\geq(4N+2)^2$. 

In our implementation, we use $L=300$ trajectories and a window length of $K=60$, for which the minimum eigenvalues of the normalized empirical information matrix remain strictly positive during training, while maintaining execution times compatible with the 15-second frequency regulation interval to be discussed in the next section.%(15 minutes at $\Delta t=15$ s). 
%Under these settings, 
%The minimum eigenvalues of the normalized empirical information matrix remain strictly positive during training, indicating sufficient excitation in practice, while maintaining execution times compatible with the 15-second frequency regulation interval to be discussed in the next section.

\color{black}
% $\bullet$ The value of $n_p$ affects the prediction accuracy. %determines the model update frequency, 
% %    which is set to 3 mins, following \cite{wang2020electric}, 
% In our simulations, with $\Delta t= 15s$, $n_p=12$, the same parameters are used over a 3 mins horizon for prediction, which yields reasonably accurate results, aligning with the observations in ~\cite{wang2020electric}. It should be noted that the estimation accuracy of Algorithm 1 may decrease over time due to, %\color{blue} 
% EVs transitioning between different states as a result of both natural dynamics, control actions, and their traveling behaviors. %\color{red}for example, uncertainties from user traveling behaviors \color{black}. 
% A moving window is employed to update model parameters every 3 minutes during frequency regulation, as detailed in the next section.  \\ %A moving window will be adopted to periodically update model parameters in the frequency regulation as will be elaborated in the next section. \\%\color{blue} The feasibility of using a time step of $\Delta t = 15$ s and keeping model parameters constant over a 3-min interval has been validated in~\cite{wang2020electric}.\color{black}\\ %$n_p\Delta t = 3 \text{ mins}$ 
%    in our simulations.
%\item %\color{red} 
\noindent$\bullet$ Due to the eSSM structure, the computational burden of parameter estimation depends mainly on the number of SOC bins $N$ that determines the system’s dimension and matrix sizes in (\ref{eq:parameters}) and dataset size ($L$, $K$), rather than the number of connected EVs $N_{EV}$. For any fixed $N$, the computational burden remains essentially constant as the EV population grows. %The execution time of the algorithm mainly depends on the number of states $N$ and dataset size ($L$, $K$).  %(e.g., the number of trajectories $L$ and the length of each trajectory $K$). 
%Specifically, $N$ determines the system’s dimension and matrix sizes in (\ref{eq:parameters}) while $L$ and $K$ determine the amount of training data. 
In our simulations,
$N=3$ offers a good balance between accuracy and efficiency.  %computational time, enabling real-time control. %Although $N$ is small, it provides sufficient accuracy while keeping computation fast enough for real-time control. 
%For example, %The number of states $N_s$ determines the dimension of the system and thus the size of the matrices in the parameter set....In our simulations, we choose $N_s=2$. 
%which seems to be a good trade-off between \color{blue}computational complexity and estimation accuracy. \color{red}...
%Another parameter is the size of the dataset. For example, \color{blue}when we collect 200 trajectories 
When $N=3$, $L=300$,  %trajectories and a time window length %and select the time window length 
$K=60$, the average estimation time is %\color{blue} 
4.6776s,  well within the 15 s frequency regulation interval to be discussed in the next section.  

\noindent$\bullet$ Algorithm 1 estimates future power outputs and flexibility solely from aggregated power measurements $\bm{Y}^l \triangleq \{P(k-K), P(k-K+1), \cdots, P(k)\}$  and the broadcast control inputs $\bm{U}^l \triangleq \{\bm{u}(k-K), \bm{u}(k-K+1), \cdots, \bm{u}(k-1)\}$. As a result, no private user information is exposed at the modeling or prediction stage. Hence, it prevents leakage
of any private user information at the modeling and prediction stage. On the other hand, the identified model is input–output equivalent only up to a similarity transformation (see (\ref{eq:similarity transformation})). Consequently, the internal state representation is not unique, and individual EV SOC trajectories cannot be reconstructed from the estimated model, providing an additional layer of privacy protection. %\color{red} Check, complete and improve it.
\color{black}

%\color{red} explain that it can be done in a moving window fashion, may add a figure. Add remarks to give the parameter values like L, k, sampling time, ... \color{black}
%\color{red} A new section is needed for control \color{black}
% \begin{figure}[t]
% \centering
% \includegraphics[width=0.9\columnwidth]{figs/overallframework.pdf}
% \caption{Framework of modeling and control.}
% \label{fig:overallframework}
% % \vspace{-0.6cm}
% \end{figure}

% \begin{figure}[t]
% \centering
% \includegraphics[width=1.0\columnwidth]{figs/MPC.pdf}
% \caption{Framework of model Predictive control.}
% \label{fig:MPC}
% % \vspace{-0.6cm}
% \end{figure}
\vspace{-0.2cm}
\section{Frequency Regulation}
% \color{blue}delete the frequency regulation part?\color{black}
%\color{red}
Once the model parameter is estimated, %....(a small paragraph for connection)\color{blue}
the bHMM can be applied to predict the power outputs and flexibility of the aggregated EVs, based on which we can design a %model predictive control (MPC) 
MPC strategy for frequency regulation. %service. 
\vspace{-0.3cm}
\subsection{Framework of Frequency Regulation with bHMM}
\begin{figure}[t]
\centering
\includegraphics[width=0.65\columnwidth]{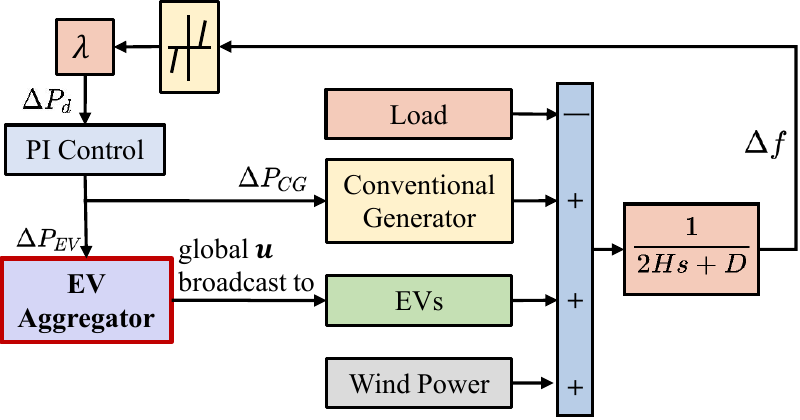}
% \vspace{-0.2cm}
\vspace{-0.5em}
\caption{The frequency regulation framework with EVs.} %\color{red} exactly the same?\color{blue}\cite{wang2020electric}}
\label{fig:FrequencyRegulation}
\vspace{-0.3em}
\end{figure}

\vspace{-0.1cm}
In this paper, we adopt the frequency regulation framework %of frequency regulation \color{red} 
from \cite{wang2019state}\color{black}. %is depicted 
As shown in Fig. \ref{fig:FrequencyRegulation}, %\color{red} explain these notations \color{black}
% the power output variation of the conventional generator (CG) $\Delta P_{CG}$ and the power variation of load and wind power $\Delta P_{d}$ change linearly, 
the frequency deviation $\Delta f(k)$ results from power imbalance %the power imbalance 
between generation and load.  %imbalance between wind power and load power. %between load power and generation power, and it can be 
%It can be calculated by an analytical method in \cite{wang2019state}. 
% The frequency regulation is firstly regulated by the load and CGs. 
% And then the EVA and CGs are utilized to implement the secondary frequency control. 
The total demanded regulation power $\Delta P_{d}$ can be obtained by the proportional-integral (PI) controller: %with \eqref{eq:dP_d}.
\vspace{-0.1cm}
\begin{equation}
    {\footnotesize
        \Delta P_{d}\left(k\right) = \begin{cases}
        \lambda\left(k\right) \cdot \left(\Delta f\left(k\right) - \Delta f_{\varepsilon}\right), & |\Delta f\left(k\right)| > \Delta f_{\varepsilon} \\
        0, & |\Delta f\left(k\right)| \leq \Delta f_{\varepsilon}
    \end{cases}
    }
    \label{eq:dP_d}
\end{equation}
where $\left[-\Delta f_{\varepsilon}, \Delta f_{\varepsilon}\right]$ is the dead band of frequency regulation, and $\lambda (k)$ is the regulation bias factor, updated via a bisection method at each time step \cite{wang2019state}. %It can be calculated by a bisection method every time step \cite{wang2019state}.
 
% Upon receiving control commands, conventional generators (CGs) respond according to their rated ramping limits, while EVs can adjust power output instantaneously. Therefore, EVs are set as the primary source for compensating power imbalances. 
Given $\Delta P_d(k)$, EVs are dispatched as the primary resource to provide the fast regulation component, limited by the available flexibility $\Delta P_{EV}(k)$ estimated by Algorithm~\ref{alg:EM}, while conventional generators (CGs) serve as a secondary resource to compensate any residual imbalance $\Delta P_{CG}(k)$ subject to their ramp-rate limits.
% The flexibility estimated by Algorithm~\ref{alg:EM} is compared with the power imbalance $\Delta P_{d}$: if the available flexibility exceeds $\Delta P_{d}$, the entire imbalance is allocated to EVs ($\Delta P_{EV}$); otherwise, the remaining portion is supplied by CGs ($\Delta P_{CG}$).
%can dispatch $\Delta P_{d}$ to EVs ($\Delta P_{EV}$) and CGs ($\Delta P_{CG}$) with the dispatch strategy in \cite{wang2019state}. \color{blue} $\Delta P_{d}$ is is first balanced by the aggregated EVs, and any remaining unbalanced power is subsequently compensated by the CGs. \color{blue} With the estimated power output and flexibility of EVs derived from the Broad Learning System (BLS) model, $\Delta P_d$ can be dynamically partitioned between EVs and CGs in real time, ensuring fast response and optimal use of available resources. \color{red} Relate to the flexibility\color{black} %\color{red} how is the frequency regulation framework and the dispatch strategy different? In other words, what is the difference between \cite{shi2017hybrid} and \cite{wang2019state}\color{black}. 
In this paper, we focus on how to design the control signals $\bm{u}$ according to  %According to 
the given $\Delta P_{EV}$. As detailed in the next subsection, we will present a model predictive control-based strategy, leveraging the estimated aggregated EV model in the previous section. The resulting global control signals $\bm{u}$ designed without individual EV information are then broadcast to all EVs, enabling each EV to respect its own operational constraints while collectively achieving the desired aggregate power-tracking performance. %we will design global identical signals $\bm{u}$ broadcast to all EVs which ensures all individual EVs obey their own operation constraints while collectively realizing the desired aggregated control performance. %the EV aggregator can generate respective control signals $\bm{u}$ and transmit it to all EVs. 
\vspace{-0.6em}

%\color{blue}
\vspace{-0.2em}
\subsection{Model Predictive Control}
MPC is an advanced control strategy that uses a system model to predict future behavior and make control decisions \cite{kouvaritakis2016model}. 
At each step of the MPC, we are solving an optimization problem of the form:
\begin{equation}
\label{eq:MPC}
\vspace{-0.1cm}
{\footnotesize
%\color{blue}
    \begin{aligned}
    \min_{\mathbf{u}} \quad &J(\bm{u}, P) \\ = &\min_{\mathbf{u}} \bigg\{\sum\limits_{m=k+1}^{k+n_c}||P(m) - P(m)^{ref}||^{2}_{\bm{Q}} +\sum\limits_{m=k}^{k+n_c-1} || \bm{u}(m)||^{2}_{\bm{R}}\bigg\}\\
    \text{s.t.} \ \ & \eqref{eq:bHMM} \text{ is satisfied;} \\
    & 0 \leq \bm{u} \leq;\\
    & \underline{P}(m) \leq P(m) \leq \overline{P}(m).
\end{aligned}
}
\end{equation}
% \begin{equation}
% \label{eq:MPC}
% \begin{small}
% %\color{blue}
%     \begin{aligned}
%     \min_{\mathbf{u}} \quad &J(\bm{u}, P) \\ = &\min_{\mathbf{u}} \bigg\{\sum\limits_{m=k+1}^{k+n_c}\left(P(m) - P(m)^{ref}\right)^{T}\bm{Q}\left(P(m) - P(m)^{ref}\right) \\
%     & +\sum\limits_{m=k}^{k+n_c-1} \left( \bm{u}(m)^{T}\bm{R}\bm{u}(m)\right)\bigg\}\\
%     \text{s.t.} \ \ & \eqref{eq:bHMM} \text{ is satisfied；} \\
%     & 0 \leq \bm{u} \leq 1；\\
%     & $\underline{P}(m) \leq P(m) \leq \overline{P}(m)$.
% \end{aligned}
% \end{small}
% \end{equation}
where $J(\bm{u}, P)$ denotes the cost function.  $\bm{Q}$ and $\bm{R}$ %\color{red} where is R? \color{black} 
are the positive definite symmetric weighting matrices, %\color{blue} 
where $\bm{Q}=\bm{I}$ and $\bm{R}=10^{-6}\bm{I}$ are chosen to prioritize power-tracking performance.  %as the primary objective. %\color{black} 
$P(m)$ and $\underline{P}(m)$/ $\overline{P}(m)$ are the estimated aggregated power output and flexibility from Algorithm \ref{alg:EM}. In our simulations, the %prediction horizon is set to be the same as the 
control horizon is set as 15s, i.e.,  %\color{blue
$\Delta t=15s$, $n_c= 1$. \color{black} The optimization repeats at every time step.

Each element of $\bm{u}$ represents the switching probability for EVs in a corresponding state.
For instance, if $u_{1}=0.3$, it indicates that EVs with an SOC corresponding to %currently in 
state $x_{1}$ in CM should switch to IM with a probability of $30\%$. 
%\color{blue}
% Under the control structure proposed in~\cite{wang2019state}, an identical control signal as defined in~\eqref{eq:def_up}—to be detailed in Section~III—can be generated. 
Upon receiving %the global control signal 
$\bm{u}$,  %as defined in~\eqref{eq:def_up}, 
each EV determines its relevant control entry %element 
in $\bm{u}$ based on its mode %connection status 
and real-time SOC. %as in Fig.~\ref{fig:state_transition}. %The corresponding element value represents the switching probability of the single EV. 
%For each EV, 
A random variable $\alpha_{i}\sim U(0,1)$  for each EV $i$ with SOC in $j$th state \color{black} is then sampled, and the switching decision is made follows:
%\color{red} explain here how the control design $u$ will obey individual EV's own charing constraints\color{black}
%\color{red} I don't see w and v in the model \color{black}
\begin{equation}
\vspace{-0.1cm}
{\footnotesize
        \begin{cases}
        \alpha_{i} \leq {u}_{j}, &\text{switch}\\
        \alpha_{i} > {u}_{j}, &\text{stay}
    \end{cases}
    \label{eq:switch}
}
\vspace{-0.4em}
\end{equation}
%\color{blue}where $u_j$ is the control input for the $j$th state. \color{red} do you need to explain i,j? i refer to EV i?\color{black} %with the corresponding EV state. 
For EVs in fully discharged/ charged states ($x_{3N+1}/x_{3N+2}$), transitions are restricted such that they can only switch to CM/DM according to %\color{blue}
$u_{d,N+1}/u_{c,N+1}$ in ~\eqref{eq:def_up}. 
It should be noted that no control entry is designed for the EVs in FCM, so they remain in CM without violating individual operational constraints. %just keep in CS without violating the operational constraints of single EVs. 

Importantly, the control signal 
$\bm{u}$ designed in~\eqref{eq:MPC} by the aggregator requires no individual EV data. By broadcasting $\bm{u}$,  the control authority is partially decentralized to individual EVs and each EV can determine its action based on its mode and SOC, ensuring compliance with local constraints while collectively achieving the desired aggregate control performance. This partially decentralized implementation reduces communication overhead and preserves user privacy without compromising system-level dispatch performance. From the grid’s perspective, the EV fleet is characterized solely by its aggregate external behavior such as adjustable capacity and dynamic response, which is explicitly modeled and optimized in \eqref{eq:MPC}, without requiring access to individual SOCs. %In this way, individual SOCs are used only for local decision-making, ensuring compliance with EV-level operational constraints, while the overall EV population collectively delivers the desired aggregate response for frequency regulation. This partially decentralized implementation significantly reduces communication overhead and preserves user privacy without affecting system-level dispatch performance. From the perspective of system dispatch and frequency regulation, the EV cluster is characterized solely by its aggregated external behavior, such as adjustable capacity and dynamic response, rather than individual EV SOCs. These aggregated characteristics are explicitly modeled and optimized by the aggregator in the control design of \eqref{eq:MPC}, which requires no individual EV data.\color{black}
% From the perspective of system operation, the EV cluster is solely by its aggregated external behavior, such as adjustable capacity  and dynamic response, rather than individual EV SOCs. \color{black}
%It is worth noting that $\bm{u}$ designed in (\ref{eq:MPC}) by the aggregator doesn't require any individual EV information, while the broadcasting $\bm{Pu}$ for each individual EV making decisions based on their own mode and SOC  ensures that  each EV obeys their own operational constraints while  collectively realizing the desired aggregate control performance. 
%This probabilistic mechanism ensures that the aggregate charging behavior follows the global control signal, while the individual EVs autonomously respect their own physical and operational constraints. Broadcasting global signals also significantly reduces communication overhead. %, such battery capacity limits and connection availability. For instance, an EV in FCS will remain charged regardless of the control command.
%Under this structure, the control authority is decentralized, allowing each EV to make local decisions that collectively realize the desired aggregate control performance. 
\begin{figure}[t]
    \centering
    \includegraphics[width=0.8\linewidth]{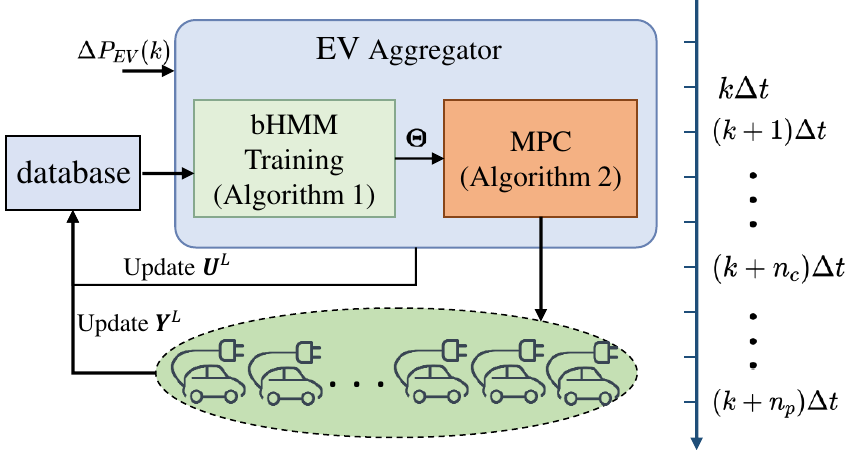}
    % \vspace{-0.2cm}
    \caption{The proposed data-driven bHMM modeling and control framework.}
    \label{fig:alg2}
    % \vspace{-0.3em}
\end{figure}
\color{black}
%\color{red}
%\color{blue}where $\bm{u}_m$ is the EV's corresponding control signal element. \color{black}

%\color{red} need to explain u is designed without the need of knowing individual EV state and how EV respond to the control command to make sure their own charging constraints are obeyed. \color{black}
The proposed data-driven privacy-preserving modeling and control of aggregated EVs %algorithm of 
for frequency regulation %in with aggregated EVs and CGs 
is summarized in Algorithm~\ref{alg:estimation_control} and illustrated in Fig. \ref{fig:alg2}.  Given historical aggregated data with $L-1$ past-day trajectories and an initial parameter set $\bm{\Theta}^{(0)}$, the bHMM is updated every 3 mins ($n_p\Delta t = 3~\text{mins}$) using new measurements and control inputs (Step 1: E1–E2), ensuring adaptability to system dynamics. %\color{blue}
Within each 3-minute interval, to mitigate risks due to model uncertainty, the new observed EV power output $P_{EV}^{real}$ is compared with the prediction value. If the prediction error exceeds 5\%, the model is updated immediately with the most recent data. \color{black} Step 2 performs real-time frequency regulation via MPC (Steps C1–C4), %including control optimization, probabilistic switching, and feedback. These control steps are executed every 15 s, regardless of model updates.Assume we have historical data set with %\color{red}
%$L-1$ past day trajectories, and initial parameter set $\Theta^{(0)}$. %...\color{red} I think you need an initial parameter set and an initial estimate of flexibility, also parameter $n_p$, $n_c$. These either inputs or parameters should be made clearer in Algorithm 2\color{black}
%generated by the  Step 1 and Step 2 represent the process of calculating required power adjustments for aggregated EVs and CGs according to the frequency deviation. 
%\color{blue} 
%Every 3 minutes ($n_p\Delta t = 3~\text{mins}$), the BLS model parameters are updated by considering the newly collected measurement and control data, following the training procedure described in Step 1 (i.e., Step E1 – Step E2 of Algorithm~\ref{alg:estimation_control}). This ensures the data-driven model remains adaptive to the evolving system dynamics. Step 2 includes real-time frequency regulation using MPC, 
%where Step C1 – Step C4 
which involves computing the optimal control, executing probabilistic EV switching, and collecting real-time power outputs from EVs under the proposed control. %of the aggregated EV response.  
%These control steps are executed every 15 s ($\Delta t = 15$), regardless of model updates. %These control steps are executed at every 15-second interval ($\Delta t = 15$ s), regardless of whether model updates occur. 
%\color{red}TBD\color{black}
%Every 3 mins ($n_p\Delta t=3 \text{mins}$), the BLS model parameters are updated with new measurement and control data following the training Step 3 (Step E1 - Step E2)....... %\color{red} I don't see E1 E2 in the algorithm \color{black} 
%Step C1  - Step C3 are for real-time MPC control. %\color{red}How is $n_k$ and K related? \color{black}\color{black}

\begin{algorithm}[t]
\LinesNotNumbered
\footnotesize
\caption{The Data-Driven Privacy-Preserving Modeling and Control of Aggregated EVs for Frequency Regulation}
\SetAlgoNlRelativeSize{-0.5}
\label{alg:estimation_control}
\KwIn{A dataset $\bm{\Phi}$ with $L$ independent trajectories (including 
$L-1$ historical and real-time trajectories of control inputs, aggregated power outputs and flexibility); %input signals and aggregated power\color{blue} power trajectories and flexibility\color{red} flexibility?\color{black}; 
initial parameter set
$\bm{\Theta}^{(0)}$; 
%The real-time frequency deviation $\Delta f(k)$} 
the real-time demanded regulation power $\Delta P_d(k+1)$; model update interval $n_p$; %\color{red}conventional generator ramp rate $r_g$...\color{blue} 
conventional generator ramp rate $r_g$ and its maximum/minimum power output $P_{CG}^{\max}$/$P_{CG}^{\min}$ \color{black}} 
\KwOut{Real-time power dispatch decision for EVs $\bm{u}(k)$ and regulation power for CGs $\Delta P_{CG}(k+1)$}
Set the initial mean absolute percentage error (MAPE) between the reference power and power output $Err_{\text{p}}=0$.\color{black} \\
\For{$k=1,2,\dots$}{
\textbf{Step 1.} Model Update and Prediction\;

\If{$k \bmod n_p = 0$ or $Err_p>5\%$}{
    \textbf{Step E1.} Move the sliding window and add latest measurements $\{P(k-n_p),\cdots,P(k)\}$ and control signals $\{\bm{u}(k-n_p-1),\cdots,\bm{u}(k-1)\}$ to dataset $\bm{\Phi}$ (see Fig.~\ref{fig:MPC});\;
    
    \textbf{Step E2.} Use $\bm{\Phi}(k)$ as inputs and apply Algorithm~\ref{alg:EM} to update the bHMM parameters $\bm{\Theta}$;\;
}
    \hspace{1.6em}\textbf{Step E3.} Use the current model parameters $\bm{\Theta}$ to predict the aggregated power outputs and flexibility $\scriptsize \bm{y}(k+1)=[P(k+1), \overline{P}(k+1), \underline{P}(k+1)]^T$\;.
    % If $k \bmod n_p \neq 0$, compare the actual EV power output $P^{\text{real}}_{EV}(k+1)$ with the prediction. If the mean absolute percentage error (MAPE) exceeds 5\%, update the dataset with the newly recorded data and proceed to Step E1. \color{black} %\color{red} k or k+1?\color{black}.
    
    \textbf{Step 2.} Power Dispatch and Frequency Regulation\; 
    
    \hspace{1.6em}\textbf{Step C1.} Compare the flexibility $\overline{P}(k+1)$ and $\underline{P}(k+1)$ estimated by Algorithm \ref{alg:EM} with 
    the power imbalance $\Delta P_{d}(k+1)$ and determine the dispatch:
    
 \If{$\scriptsize \Delta P_d(k+1) > 0$}{
    \eIf{$\scriptsize \overline{P}(k+1) - P(k+1) > \Delta P_d(k+1)$}{
      $\scriptsize \Delta P_{EV}(k+1) = \Delta P_d(k+1)$;\;
    }{
      $\scriptsize \Delta P_{EV}(k+1) = \overline{P}(k+1) - P(k+1)$,
      
      $\begin{aligned}
        \Delta P_{CG}(k+1) = \min \big(
        & P_{CG}^{\max} - P_{CG}(k),\ r_g \cdot \Delta t, \\
        & \Delta P_d(k+1) - \Delta P_{EV}(k+1) \big).
    \end{aligned}
$\;
    }
  }
  \Else{
    \eIf{$\scriptsize \underline{P}(k+1) - P(k+1) < \Delta P_d(k+1)$}{
      $\scriptsize \Delta P_{EV}(k+1) = \Delta P_d(k+1)$;\;
    }{
      $\scriptsize \Delta P_{EV}(k+1) = \underline{P}(k+1) - P(k+1)$,
      
       $\begin{aligned}
        \Delta P_{CG}(k+1) = \max \big(
        & P_{CG}^{\min} - P_{CG}(k),\ -r_g \cdot \Delta t, \\
        & \Delta P_d(k+1) - \Delta P_{EV}(k+1) \big).
    \end{aligned}
$\;
    }
  }
    \hspace{1.6em}\textbf{Step C2.}  Set $\scriptsize P^{\text{ref}}(k+1)= \Delta P_{EV}(k+1)+P(k+1)$ in \eqref{eq:MPC} and solve the MPC problem with $\bm{\Theta}$ for optimal control commands $\bm{u}(k)$; broadcast $\bm{u}(k)$ to all EVs;
    
    \hspace{1.6em}\textbf{Step C3.} For each EV $i$, %a random variable 
    sample $\alpha_i\sim ~U(0,1)$ %is generated, 
    and apply probabilistic switching rule \eqref{eq:switch}
    ; %to make the switching decisions.
    
    \hspace{1.6em}\textbf{Step C4.} Aggregator records the actual power output of EVs $P^{\text{real}}_{EV}(k+1)$ under $\bm{u}(k)$, %\color{blue} 
    and updates the tracking performance  MAPE  $Err_p= |(P^{\text{ref}}(k+1)-P_{EV}^{\text{real}}(k+1))/P^{\text{ref}}(k+1)|$ %\color{red} do you need index k? 
}
\end{algorithm}
\textbf{Remarks:}\\
%\begin{enumerate}
    %\item  
    $\bullet$ $n_p$ determines %controls 
    %\color{blue} 
    the model parameter update frequency, %\color{black}
    % the prediction horizon, 
    which is set to 3 mins, following \cite{wang2020electric}, with $\Delta t= 15s$, $n_p=12$. %$n_p\Delta t = 3 \text{ mins}$ 
   % in our simulations. 
   Model updates occur when $k \bmod n_p=0$. %When $k \bmod n_p=0$, the horizon reaches the end and the model parameters are updated. 
   A larger $n_p$ reduces the communication burden for measurement and control updates but may increase estimation errors. 3 mins provides a good trade-off between accuracy and communication cost. \\
   % \item 
     $\bullet$ The control %horizon and 
    interval is set to 15s following~\cite{wang2020electric}, i.e., the control is implemented at each time step. When no parameter updates are needed ($k \bmod n_p \not= 0$), the average MPC computation time is %\color{blue} 
    0.3301s. When the model parameters need updates ($k \bmod n_p = 0$), as mentioned in Remark 2 of Algorithm \ref{alg:EM}, it takes about %\color{blue}
    4.6776s, bringing the total average time (including MPC) to 
    5.0077s—well under the control interval of 15s, ensuring Algorithm \ref{alg:estimation_control} is feasible for real-time implementation. %Combined with the MPC time, the total average time of Algorithm \ref{alg:estimation_control} is 12.59 s- still under the 13-s control interval, ensuring the feasibility for online implmenetation. %below 15s, making the proposed method feasible for real-time control. %while the control %horizon and 
\section{Numerical Validation}\label{section:numerical validation}
%\color{red}
In this section, we test the proposed data-driven privacy-preserving modeling and control algorithm of aggregated EVs for frequency regulation. %BLS in Section III. \color{blue} 
First, Algorithm \ref{alg:EM} is tested by comparing the predicted flexibility and aggregated power outputs from the bHMM with those from the eSSM \cite{liu2025extended} and the benchmark individual modeling method (IMM) \cite{zhang2016evaluation}. The IMM, which sums the aggregated power output and flexibility of all EVs modeled by \eqref{eq:SOC_variation}, serves as the reference benchmark.
%IMM computes the aggregated power output and flexibility by summing across all EVs %It obtained the power output and flexibility by calculating the summation of every individual EV 
%modeled by (\ref{eq:SOC_variation}), which is treated as the true benchmark model. 
%\color{red} Did you explain IMM or use the term IMM before? I think you use Individual EV in section II.A\color{black}. 
Next, we assess the frequency regulation performance of Algorithm~\ref{alg:estimation_control}, comparing its results to %those achieved with 
eSSM-based control and FL-based control, the latter representing an emerging privacy-preserving approach. %\color{red} also FL-based one. 
All experiments were conducted on a desktop with an Intel Core i7-10700 @ 2.90 GHz and 16 GB RAM (no discrete GPU). \color{black} %with the frequency regulation control result based on eSSM. 
%It is worth noting that eSSM requires all detailed individual information (e.g., the SOC values of all EVs, the charging and discharging power), while the proposed BLS does not. %fromoutput powertrajectories are compared with eSSM and IMM. Given the prediction results, the control performance of the proposed control method is also tested and validated.
% The predicted aggregated power trajectory will be compared with the model-based SSM model\cite{wang2019state} and baseline the individual modeling method (IMM)\cite{zhang2016evaluation}. \color{black} 
\begin{table}[!ht]
%\color{blue}
\caption{Rated Charging/Discharging Power and Proportions of EVs}
\vspace{-0.1cm}
\label{tab:charge_power}
\centering
\renewcommand{\arraystretch}{1.2}
\setlength\tabcolsep{8pt} % 调整列距
\begin{tabular}{l l l l l l }
\hline
\hline
$P_{c}/P_{d} (kW)$ & $6.2$ & $7.2$ & $9.6$ & $11.5$ & $19.2$\\
\hline
Proportion(\%) & $85.25$ & $13.80$ & $0.21$ & $0.53$ & $0.21$\\
\midrule
\specialrule{0em}{1pt}{1pt}
\midrule
\end{tabular}
% \vspace{-1em}
\end{table}

\begin{table}[htb]
\vspace{-0.1cm}
\caption{Characteristic Parameters of Aggregated EVs}
\vspace{-0.2cm}
\label{tab:characteristic_parameters}
\centering
\renewcommand{\arraystretch}{1.2}
\setlength\tabcolsep{8pt} % 调整列距
\begin{tabular}{l l l}
\hline
\hline
Parameter & Description & Value$^{*}$ \\
\hline
% $P_{c}/P_{d}$ & Charging/Discharging Power (kW) & U(5.0, 7.0) \\
$\eta_{c}/\eta_{d}$ & Charging/Discharging Efficiency & U(0.88, 0.95) \\
Q & Battery Capacity (kWh) & U(20.0, 30.0) \\
\midrule
\specialrule{0em}{1pt}{1pt}
\midrule
\end{tabular}
\begin{tablenotes}
\item * $U(a, b)$ denotes a uniform distribution with variation range $[a, b]$.
\end{tablenotes}
% \vspace{-0.1cm}
\end{table}
% \vspace{-0.5cm}
\subsection{Simulation Parameters}
%\color{blue}%It is assumed that there are 

We consider an aggregator controlling $10,000$ EVs for frequency regulation. Based on real-world data on charging power and station distribution in Québec \cite{donneesquebec_bornes_2015}, the simulated fleet includes both AC Level 1 and Level 2 chargers, allocated according to Table ~\ref{tab:charge_power}.
%Assuming 10,000 vehicles are controlled by an aggregator for frequency regulation. \color{blue}Drawing on real-world data for charging power and station distribution in Québec~\cite{donneesquebec_bornes_2015}, we simulate 10, 000 EVs with AC Level~1 and Level~2 charging levels allocated in the same proportions as Table~\ref{tab:charge_power}. %\color{red} are you trying to say parameters in Table II are obtained from real-life Quebec data? \color{black}%under the centralized control of an EV aggregator for frequency regulation. 
Similar to the approach in \cite{wang2019state}, EVs' charging/discharging efficiencies and capacities follow uniform distributions (TABLE \ref{tab:characteristic_parameters}) and traveling parameters follow normal distributions (TABLE \ref{tab:traveling_parameters}). %from \cite{wang2019state}. %Individual EV parameters are randomly sampled from these distributions. 
\color{black}
Each EV’s rated charging power and efficiency are assumed equal to its discharging counterparts. Upon connection to the grid, the EVs start being charged until $S_{\max}$, unless interrupted by aggregator control. 
\begin{table}
\vspace{-0.3em}
\caption{Traveling Parameters of Aggregated EVs}
\vspace{-0.4em}
\label{tab:traveling_parameters}
\centering
\renewcommand{\arraystretch}{1.0}
\setlength\tabcolsep{6pt}
\begin{threeparttable}
\begin{tabularx}{\linewidth}{l l l}
\hline
\hline
Parameter & Description & Value$^{*}$ \\
\midrule
$S_{s,i}$ & Start Charging SOC & $N(0.3, 0.05) \in [0.2, 0.4]$ \\
$S_{d,i}$ & Demanded SOC for Travel & $N(0.8, 0.03) \in [0.7, 0.9]$ \\
$t_{s,i}$ & Start Charging Time (h) & 
\begin{tabular}[t]{@{}l@{}}
$N(-6.5, 3.4) \in [0.0, 5.5]$,\\
$N(17.5, 3.4) \in [5.5, 24.0]$
\end{tabular} \\
$t_{f,i}$ & Finish Charging Time (h) & 
\begin{tabular}[t]{@{}l@{}}
$N(8.9, 3.4) \in [0.0, 20.9]$,\\
$N(32.9, 3.4) \in [20.9, 24.0]$
\end{tabular} \\
$S_{\min}/S_{\max}$ & Min/Max SOC Value & 0.0 / 1.0 \\
\hline
\hline
\end{tabularx}
\begin{tablenotes}
\item[*] $N(\mu, \sigma)$ represents the normal distribution, where $\mu$ is the mean, $\sigma$ is the standard deviation, and $[\alpha, \beta]$ is the variation range.
\end{tablenotes}
\end{threeparttable}
\end{table}
% \vspace{-0.3cm}
%EVs' rated charging power and charging efficiency are assumed to be equal to their rated discharging power and discharging efficiency. 
%Besides, after an EV plug into the grid, it is assumed to start charging until disturbed by any control commands from the EV aggregator. Parameters of each EV are randomly drawn from the given distributions above.

%The individual modeling method (IMM)\cite{zhang2016evaluation}, commonly used as a benchmark %baseline to test the accuracy of other modeling methods 
%\cite{lam2015capacity}, computes the aggregated power output and flexibility by summing across all EVs %It obtained the power output and flexibility by calculating the summation of every individual EV 
%modeled by (\ref{eq:SOC_variation}). The eSSM is also simulated for comparison. %modeling method is shown as a comparison.
%\color{red} Table I, II should be before Fig. 7. They should appear according to their orders in the text\color{black}

%TABLE. \ref{tab:para_freq} shows the parameters for frequency regulation. Fig. \ref{fig:wind_load} depicts the profiles of the wind and load power. We can observe that wind fluctuation is the primary cause of power imbalance. %Simulation 
%The time interval $\Delta t=15s$. %is set as 15s. 
%For eSSM and BLS, the number of SOC intervals is set to be 2 (i.e. $N_s=2$).
%For the MPC part, the prediction horizon $n_{p}$ is 3 min (i.e. $n_{p}=12$) and control horizon $n_{c}$ is 15 s (i.e. $n_{c}=1$).
% During the frequency regulation, 
% \vspace{-0.1cm}
\subsection{The Performance of the Proposed Data-Driven bHMM in Estimating the  Power and Flexibility} %Bilinear System Model}
%Based on the traveling behaviors of aggregated EVs, we collect 50 independent daily power trajectories of a group of aggregated EVs (i.e. $L=50$) with 60 data points ($K=60$).
%Information is updated every $T_{p}=n_p\Delta t=3min$ period. \color{red} TBD: To see if these be given later \color{black}
%\color{blue} Collecting $200$ trajectories\color{red}.. 
We collect 300 trajectories ($L=300$) by sampling characteristic and traveling parameters from TABLE \ref{tab:charge_power}, \ref{tab:characteristic_parameters} and \ref{tab:traveling_parameters}. %\color{black}, each with 60 data points ($K=60$). %\color{blue} %Start from the beginning, 
A sliding time window of 15 mins ($K=60$) is used to estimate the parameters $\bm{\Theta}$ and predict the aggregated power output and flexibility. The model parameters are updated every 3 mins ($\Delta t= 15s$, $n_p=12$).

% \vspace{-0.3em}
% \begin{figure}
%     \centering
%     \includegraphics[width=0.95\linewidth]{figs/pre_notraj_update.pdf}
%     \caption{\color{blue}Prediction results of different modeling methods when no controls are present.} %\color{red} mark upper and lower bounds \color{black}}
%     \label{fig:pre_noinput}
%     \vspace{-0.5em}
% \end{figure}

% When there are no control commands ($\bm{u}=\bm{0}$ and no need to estimate $\bm{V}$ in \eqref{eq:parameters}), the prediction results of different modeling methods are shown in Fig. \ref{fig:pre_noinput}. 
% Fig. \ref{fig:pre_noinput} (a) presents the trajectories of the total power output of aggregated EVs, while Fig. \ref{fig:pre_noinput} (b) shows the predicted flexibility. %trajectories %results 
%of the flexibility of aggregated EVs. 
% Compared to the eSSM, the proposed data-driven bHMM achieves similar accuracy for both the power output and the flexibility under natural state transitions. 
 % similar aaccurately estimate the total power output and  flexibility trajectories of aggregated EVs  %The data-driven prediction model can accurately model the aggregated EVs 
%when there are only natural state transitions.

\begin{figure}
    \centering
    \includegraphics[width=0.98\linewidth]{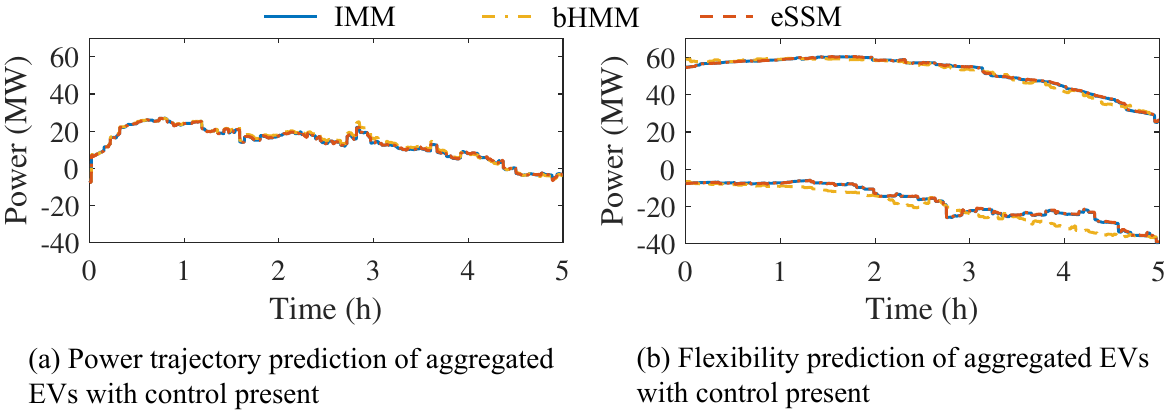}
    \vspace{-0.3cm}
    \caption{
    Prediction results of different modeling methods with control present.\color{black}}
    % \vspace{-0.5em}
    \label{fig:pre_input}
\end{figure}

%\color{blue}
% When control is applied, the parameter set to be estimated gets larger ($\bm{V}$ is no longer zero in \eqref{eq:parameters}). %Is it the only difference? \color{blue} when there are no control, we do not need to estimate V anymore\color{black}). 
Given random control inputs generated every minute (every 4 time steps), prediction results are presented in Fig. \ref{fig:pre_input}. Fig. \ref{fig:pre_input} (a) presents the trajectories of the total power output of aggregated EVs, while Fig.~\ref{fig:pre_input} (b) shows the predicted flexibility. 
% These results show that the proposed %ata-driven BLS modeling 
% method  accurately %can give accurate 
% predicticts both power outputs and flexibility of aggregated EVs when random control signals %to EVs when control signals 
% are introduced.  
The results demonstrate that the proposed method can accurately predict the aggregate EV power response under control actions. Although the flexibility estimation accuracy is slightly lower than that of the eSSM benchmark, the performance remains satisfactory given that no individual EV information is available in our privacy-preserving framework.

%\color{blue}

\textit{Sensitivity to the number of EVs ($N_{EV}$):}\color{black} By varying %the number of EVs 
$N_{EV}$ while keeping the proportion of different charging levels fixed, the estimation results over 5 hours are presented in TABLE. \ref{tab:error_Nev}. %\color{red} check table number. I think table V hasn't been commented. \color{black} 
As $N_{EV}$ increases, the estimation error decreases, demonstrating the scalability of both the proposed approach %\color{blue} 
and eSSM method. However, when $N_{EV}$ is small, the accuracy of both models declines due to stochastic fluctuations in smaller EV populations, which lead to greater deviations from the modeled aggregate dynamics—consistent with observations in \cite{wang2019state}. %It should be noted that the accuracy of both eSSM and bHMM decreases when $N_{EV}$ is small as stochastic fluctuations for samll EV polutition cause larger deviations from the modeled aggregate dynamics. Similar observations were obtained in \ref{wang2019state}. The eSSM model becomes a more accurate and stable representation of fleet behaviror as the number of EVs increases. % due to the fact that eSSM is for aggregated EVs. The matrices are more accurate and stable representations of EV the model formulation r \color{red}It should be noted that both eSSM and bHMM .....\color{blue} This is because the probalistic control mechanism and model formulation rely on enough EV population. 
%\color{red}
% For a large number of EVs, the bHMM consistently achieves estimation accuracy comparable to that of the eSSM when the number of EVs exceeds 5,000 and even surpasses it when $N_{EV}$ reaches 80,000. 
The applicable range of the proposed method should be greater than $5,000$. %\color{red} if $<10\%$ is treated as good. But any justification for this value? 
It is worth noting that, unlike eSSM, the proposed data-driven bHMM requires no  %eSSM requires all detailed 
detailed individual EV data.  %(e.g., the SOC, the charging and discharging power and efficiency).  \color{black}

\textit{Sensitivity to the number of SOC state interval $N$:} By varying $N$, %the number of SOC state intervals, 
TABLE.~\ref{tab:error_Ns} reports the average execution time and the corresponding estimation accuracy. As $N$ increases, the model achieves improved accuracy by providing a finer discretization of the SOC distribution and thus a more detailed representation of the fleet dynamics. However, the computational complexity (average executive time) of both training and online control increases significantly as $N$ grows. These results explicitly demonstrate the trade-off between modeling accuracy and computational scalability, and justify the selection of $N=3$ as a practical compromise for real-time frequency regulation.\color{black}

\begin{table}[t]
\caption{%\color{red} should you use MAPE for consistency\color{black}
MAPE (\%) of Power Output with Different EV Populations $N_{EV}$}
% \vspace{-0.2cm}
\label{tab:error_Nev}
\centering
\renewcommand{\arraystretch}{1.1}
\begin{tabular*}{\columnwidth}{@{\extracolsep{\fill}}llllll}
\hline\hline
$N_{EV}$ & 200 & 1,000 & 5,000 & 10,000 & 80,000 \\
\hline
% bHMM & 33.72 & 17.86 & 6.384 & 3.478 & 1.114 \\
% eSSM & 13.71 & 10.17 & 4.799 & 2.900 & 1.940 \\
% bHMM & 50.40  & 15.66 & 8.71 & 4.66 & 3.25 \\
bHMM & 16.97  & 9.36 & 6.53 & 3.59 & 1.17 \\
eSSM & 11.23 & 7.53 & 4.08 & 2.13 & 1.94 \\
% bHMM & 41.65  & 13.03 & 7.96 & 3.66 & 3.07 \\
% eSSM & 13.71 & 10.17 & 4.80 & 2.90 & 1.94 \\
\hline\hline
\end{tabular*}
\end{table}
% \begin{tabularx}{\linewidth}{lXXX}
% \hline\hline
% Methods & eSSM & FL-based & bHMM \\
% \hline
% Control Error (\%) & 0 & 3.95 & 1.49 \\
% \hline\hline
% \end{tabularx}
% \vspace{-0.6em}
\begin{table}[t]
\caption{%\color{red} should you use MAPE for consistency\color{black}
Estimation Performance with Different Number of State Intervals $N$}
% \vspace{-0.2cm}
\label{tab:error_Ns}
\centering
\renewcommand{\arraystretch}{1.1}
\begin{tabular*}{\columnwidth}{@{\extracolsep{\fill}}llllll}
\hline\hline
$N$ & 2 & 3 & 5\\
\hline
% bHMM & 33.72 & 17.86 & 6.384 & 3.478 & 1.114 \\
% eSSM & 13.71 & 10.17 & 4.799 & 2.900 & 1.940 \\
Average Executive Time (s) & 0.1761 &  0.3898 & 2.3681 \\
Estimation Accuracy (\%) & 4.84   & 3.59  & 3.49 \\
\hline\hline
\end{tabular*}
\end{table}
\color{black}

\begin{table}[t]
\caption{%\color{red} should you use MAPE for consistency\color{black}
Sensitivity Analysis of different $L$ and $K$ ($N=3$)}
% \vspace{-0.2cm}
\label{tab:error_LK}
\centering
\renewcommand{\arraystretch}{1.1}
\begin{tabular*}{\columnwidth}{@{\extracolsep{\fill}}lllll}
\hline\hline
\diagbox{$K$}{$L$} & 60  & 200 & 300 & 400\\
\hline
% bHMM & 33.72 & 17.86 & 6.384 & 3.478 & 1.114 \\
% eSSM & 13.71 & 10.17 & 4.799 & 2.900 & 1.940 \\

24 (6 mins) & 7.53 & 5.90 & 4.60 & 4.54\\

% 36 (9 min) & 20.38  &  5.81 & 4.69 & \\
48 (9 mins) & 5.44  &  5.01 & 4.93  & 4.98\\
60 (12 mins) & 4.62 & 3.91 & 3.59 &  3.87\\
72 (15 mins)& 4.43 & 5.36 & 4.89 & 4.87\\
% 90 & 38.27 & &\\
\hline\hline
\end{tabular*}
\end{table}
\color{black}
% \begin{table}[t]
% \color{blue}
% \caption{%\color{red} should you use MAPE for consistency\color{black}
% Sensitivity Analysis of different $L$ and $K$ ($N=3$)}
% % \vspace{-0.2cm}
% \label{tab:error_Ns}
% \centering
% \renewcommand{\arraystretch}{1.1}
% \begin{tabular*}{\columnwidth}{@{\extracolsep{\fill}}llllll}
% \hline\hline
% $L$ & 100 & 200 & 300 & 400\\
% \hline
% % bHMM & 33.72 & 17.86 & 6.384 & 3.478 & 1.114 \\
% % eSSM & 13.71 & 10.17 & 4.799 & 2.900 & 1.940 \\

% 24 (6 min) & 2.18 &  2.67 & 2.62 & \\

% % 36 (9 min) & 2.56 &  3.09 & 2.62 & \\
% 48 (9 min) & 2.80  &  10.19 & 3.49 & \\
% 60 (12 min) & 9.49   & 3.71  & 3.77 & \\
% 72 (15 min)  & 30.62 & 3.99 & 3.96 &\\
% % 90 & 38.27 & &\\
% \hline\hline
% \end{tabular*}
% \end{table}

%\subsection{The Performance of the Proposed Data-Driven Modeling Under Non-Gaussian Assumptions}
\textit{Sensitivity to the number of trajectories $L$ and the trajectory length $K$:} TABLE.~\ref{tab:error_LK} presents the sensitivity of the estimation error to the number of trajectories $L$ and the data length $K$ when $N=3$. Overall, the results indicate that increasing $L$ generally improves estimation accuracy. 
In contrast, increasing the data length $K$ does not consistently reduce the error. While performance improves as $K$ increases from 24 to 60 in most cases, a further increase to $K=72$ often leads to no improvement or even slight degradation. Among all tested settings, $L=300$ and $K=60$ achieve the best performance. Therefore, these values are adopted in the subsequent study.

% \textit{Sensitivity to $L$ and $K$:} With $L=300$, $K=60$, varying the prediction horizon, the estimation results are listed in 

% \begin{table}[t]
% %\color{blue}
% \caption{%\color{red} should you use MAPE for consistency\color{black}
% MAPE (\%) of Power Output with Different EV Populations $N_{EV}$}
% % \vspace{-0.2cm}
% \label{tab:error_Nev}
% \centering
% \renewcommand{\arraystretch}{1.1}
% \begin{tabular*}{\columnwidth}{@{\extracolsep{\fill}}llllll}
% \hline\hline
% $n_p$ & 4 & 8 & 12 & 24 & 36 \\
% \hline
% % bHMM & 33.72 & 17.86 & 6.384 & 3.478 & 1.114 \\
% % eSSM & 13.71 & 10.17 & 4.799 & 2.900 & 1.940 \\
% % bHMM & 50.40  & 15.66 & 8.71 & 4.66 & 3.25 \\
% MAPE (\%) &   &  &  &  &  \\
% Average Executive Time (s) &  &  &  &  &  \\
% \hline\hline
% \end{tabular*}
% \end{table}
 
\textit{Sensitivity to the probability distribution of individual EV parameters:} The Gaussian assumptions about $\bm{w}$ and $\bm{v}$ in  the proposed bHMM model (\ref{eq:bHMM}) %are introduced mainly to enable tractable EM-based parameter estimation and state inference. These assumptions 
are applied at the aggregated fleet level, not on individual EV dynamics. For large EV populations, this approximation is justified by the central limit theorem. 
%For large EV fleets, aggregate deviations can be viewed as the sum of many weakly correlated disturbances. 
%By the central limit theorem, the aggregated can therefore be approximated as Gaussian, even if individual EV behaviors non-Gaussian, with the approximation improving as fleet size increases. %For large EV fleets, the aggregate deviation can be regarded as the sum of many weakly correlated random disturbances. Under such conditions, the central limit theorem suggests that the aggregated innovations can be reasonably approximated by a Gaussian distribution, even if individual EV behaviors are discrete or non-Gaussian. This approximation becomes increasingly accurate as the fleet size grows, since the aggregated SOC evolution and power response become smoother than individual EV trajectories.
To examine the robustness of the proposed model to distributional assumptions, an additional simulation was performed in which the initial SOC $S_{s,i}$ and demanded SOC $S_{d,i}$ are sampled from a uniform distribution instead of the Gaussian distributions. The results summarized in TABLE.~\ref{tab:error_distribution} show comparable prediction errors under both settings. This indicates that the proposed aggregated model is relatively insensitive to the specific distributions of individual EV parameters. %.relatively insensitive to the specific individual EVdistribution assumptions.

% Implementing the control process as will be discussed in the next section, the control results are depicted in Fig.~\ref{fig:control_uniform} with the absolute control error being 0.0584 MW (closed to 0.0544 of Gaussian distribution).

%Overall, although individual EV behaviors may follow non-Gaussian distributions, the Gaussian assumption provides a practical and sufficiently accurate approximation for aggregated EV dynamics in large-scale EV fleets.

\begin{table}[t] 
\caption{MAPE (\%) OF POWER OUTPUT UNDER DIFFERENT DISTRIBUTIONS} %\vspace{-0.15cm} 
% \small
\label{tab:error_distribution} 
\centering
\renewcommand{\arraystretch}{1.1}
\begin{tabular*}{\columnwidth}{@{\extracolsep{\fill}}llllll}
\hline\hline
$N_{EV}$ & 200 & 1,000 & 5,000 & 10,000 & 80,000 \\ 
\hline 
Gaussian & 16.97 & 9.36 & 6.53 & 3.59 & 1.17 \\ 
Uniform & 16.98 & 8.94 & 6.68 & 3.14 & 1.49 \\
% Uniform & 45.55 & 17.37 & 7.48 & 4.40 & 3.87 \\ 
\hline\hline \end{tabular*} 
% \vspace{-0.15cm} 
\end{table}

% \begin{figure}
%     \centering
%     \includegraphics[width=0.9\linewidth]{figs/freq_uniform.pdf}
%     \caption{\color{blue}Control results under uniform distributions.\color{black}}
%     \label{fig:control_uniform}
% \end{figure}

\subsection{The Performance of the Proposed Data-Driven Modeling and Frequency Regulation Algorithm} %Frequency Regulation with Bilinear System Model}

\begin{figure}[t]
    \centering
    % \vspace{-0.3em}
    \includegraphics[width=0.6\linewidth]{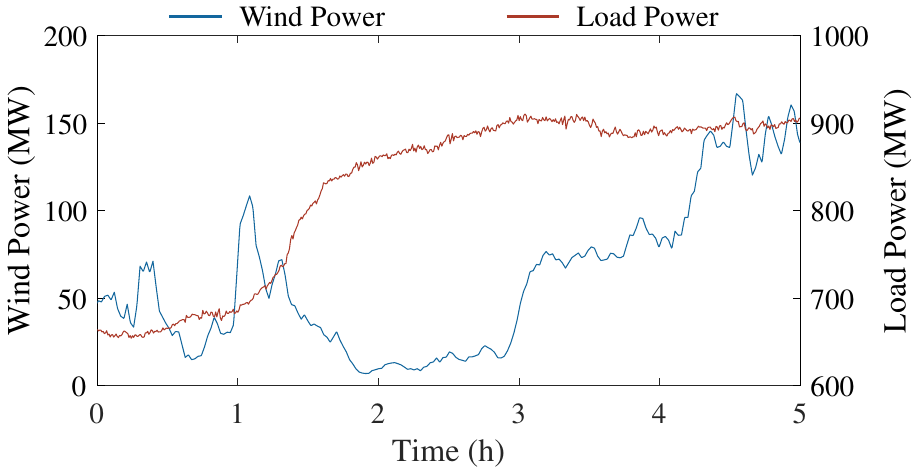}
    \vspace{-0.3cm}
    \caption{Wind and load profiles adapted from \cite{ceps_services}.}
    \label{fig:wind_load}
    \vspace{-0.5em}
\end{figure}

TABLE. \ref{tab:para_freq} shows the parameters for frequency regulation simulations, which are obtained from previous work \cite{wang2019state}.  Fig. \ref{fig:wind_load} presents real-life profiles of the wind and load power obtained from \cite{ceps_services}. %\color{red} give the source \color{black}. 
We can observe that wind fluctuation is the primary cause of power imbalance. The proposed data-driven privacy-preserving modeling and frequency regulation algorithm is implemented to maintain the frequency stability of the system. %To maintain stability frequency, the MPC is implemented every 15s. %control interval is 15s%Simulation 
%The time interval $\Delta t=15s$. %is set as 15s. 
%For eSSM and BLS, the number of SOC intervals is set to be 2 (i.e. $N_s=2$).
%For the MPC part, 
%the prediction horizon $n_{p}$ is 3 min (i.e. $n_{p}=12$) and control horizon $n_{c}$ is 15 s (i.e. $n_{c}=1$).

% During the frequency regulation, 
%To investigate the performance of the BLS model for frequency regulations, 
%\color{blue}
For comparison, we also evaluate the proposed method against the FL-based method in  \cite{qian2023federated}, which offers privacy-preserving capability. %\color{blue}
For frequency regulation, the power adjustment is firstly balanced by EVs, and CGs with ramping rates in  TABLE~\ref{tab:para_freq} are used as backup sources when EVs cannot provide sufficient flexibility.
\vspace{-0.2em}%To validate the privacy-preserving property, the FL-based control method in \cite{qian2023federated} is simulated for comparison.
% For comparison, we also 
%\color{red} I think scenario may not even be needed. Rather, need to include runtime and memory usage comparions \color{black}
% \color{blue}Two scenarios are considered : I. CGs with ramping rates in TABLE \ref{tab:para_freq}, EVs are uncontrolled resources; II. The power adjustment requirement is firstly balanced by EVs, and CGs are used as backup sources when EVs cannot provide sufficient flexibility. And in Scenario II, we consider three modeling and control methods: i. eSSM model with rule-based dispatch control method for frequency regulation~\cite{wang2019state}; %\color{red}cite \color{black} %(iii) eSSM (model-based) and MPC frequency regulation.
% ii. the control algorithm with federated learning \cite{qian2023federated};
% iii. the proposed data-driven modeling and control algorithm.
\color{black}
% \vspace{-0.3em}

\begin{table}
\caption{Parameters for Frequency Regulation}
\vspace{-0.5cm}
\label{tab:para_freq}
\renewcommand{\arraystretch}{1.0}
\centering
\setlength\tabcolsep{6pt}
\begin{threeparttable}
\begin{center}
\begin{tabular}{l l l l l l}
\hline
\hline
Parameter & Description &$\text{Value}^{*}$\\
\hline
$\Delta f_{\epsilon}$ & Maximum Frequency Deviation (Hz) & 0.1\\
H &System Inertia Constant (MW$\cdot \text{s/Hz}$) & 120 \\
D & Load Damping Coefficient (MW/Hz) & 20  \\
$r_{g}$ & Generation Ramp Rate (MW/min) & 50  \\
$P_{CG}^{\min}$/$P_{CG}^{\max}$ & Max/Min Power Output of CGs (MW) &0/500  \\
\hline
\hline
\end{tabular}
\begin{tablenotes}[flushleft]
\item * The value of $\Delta f_{\epsilon}$ derives from \cite{wang2020electric}. $D$ and $H$ comes from \cite{shi2017hybrid}. $r_{g}$ derives from \cite{meus2017applicability}. $P_{CG}^{\min}$ and $P_{CG}^{\max}$ are set to be $0/500$ MW according to \cite{wang2019state}.
% The data of the traveling parameters derive from \cite{wang2019state}.
\end{tablenotes}
\end{center}
\end{threeparttable}
\vspace{-0.5em}
\end{table}
\begin{figure}[t]
% \vspace{-0.2em}
    \centering
    \includegraphics[width=0.95\linewidth]{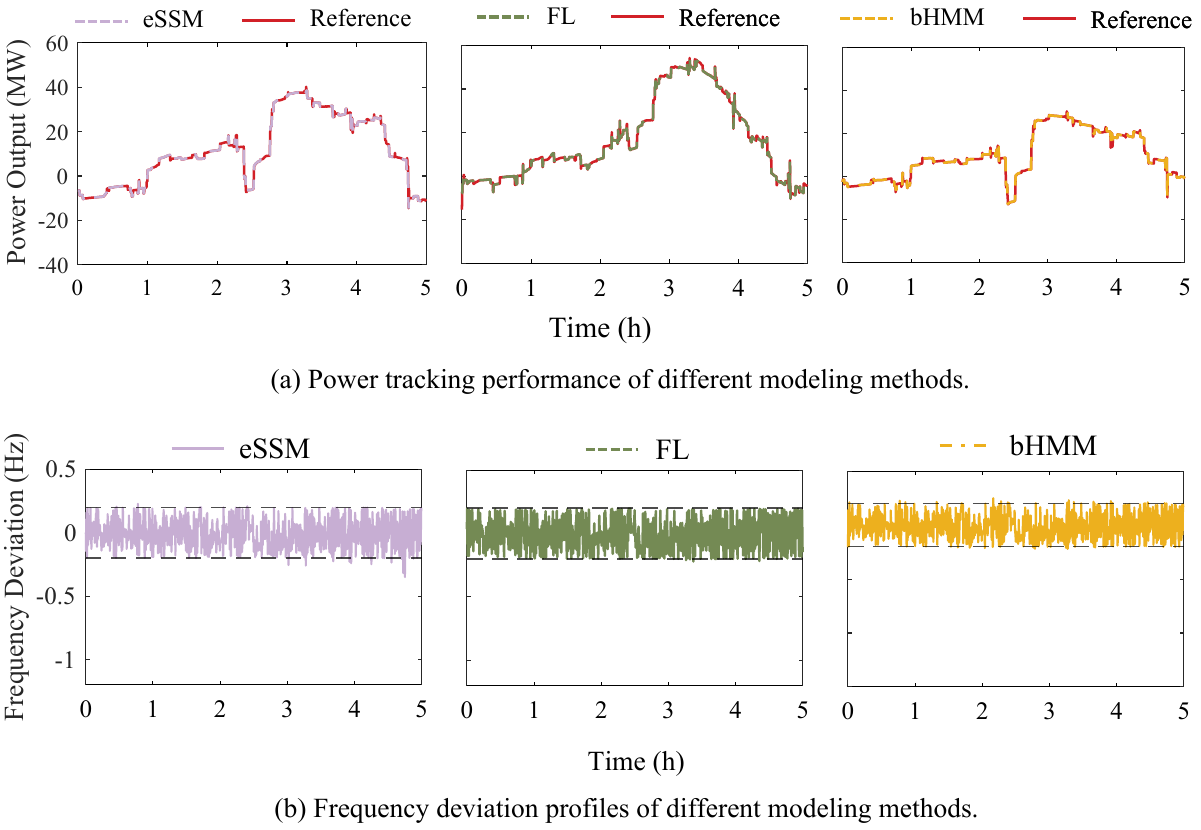}
    % \vspace{-0.3cm}
    \caption{Power tracking and frequency deviation profiles of different modeling methods.}
    % \vspace{-0.2cm}
    \label{fig:frequency_N2}
    % \vspace{-0.3em}
\end{figure}
% \vspace{-0.2cm}
% \begin{figure}
% \vspace{-0.2em}
%     \centering
%     \includegraphics[width=0.9\linewidth]{figs/control_traj.pdf}
%     \vspace{-0.3em}
%     \caption{\color{blue}Power Tracking Performance under different modeling methods in Scenario II.}
%     \vspace{-0.3em}
%     \label{fig:control_curve}
% \end{figure}

\begin{table}[!htbp]
%\color{blue}
% \vspace{-0.2cm}
\caption{
Control Performance of Different Methods}
\vspace{-0.2cm}
\label{tab:error_control}
\centering
\renewcommand{\arraystretch}{1.1}
\begin{tabularx}{\linewidth}{lccc}
\hline\hline
Methods & eSSM & FL & bHMM \\
\hline
MAPE (\%) & 0.74 & 2.45 &1.86\\
Average Communication Burden$^{*}$ & 9.67 kB &6.88 kB & 64 byte \\
Average Executive Times (s) & 0.0019 &0.1641 & 0.7199\\
\hline\hline
\end{tabularx}
\begin{tablenotes}[flushleft]
\item $*$ The communication burden represents the average amount of data transmitted per control cycle (15s) between the aggregator and EVs.
\end{tablenotes}
% \vspace{-0.5em}
\end{table}

%\color{blue}%The experiments were run on a desktop with an Intel Core i7-10700 @ 2.90 GHz and 16 GB RAM (no discrete GPU). 
Simulating the three modeling and control methods, the control and frequency regulation performance of different modeling methods are presented in Fig.~\ref{fig:frequency_N2}, where Fig.~\ref{fig:frequency_N2} (a) shows the corresponding power tracking performance and Fig.~\ref{fig:frequency_N2} (b) shows the frequency deviation profiles.
The power output control errors, average communication burden and executive times of different methods over 5 hours are summarized in TABLE.~\ref{tab:error_control}. 
%\color{blue} Corresponding power profiles of the proposed data-driven control method and eSSM are depicted in Fig. \ref{fig:traj}. %\color{red} May also show the deltaPEV? the tracking performance?  \color{red} add Table number. Also, where are there only two?\color{black} 
%\color{black}
As we can observe, the proposed method achieves power-tracking performance comparable to the eSSM approach and superior to the FL-based method, %despite not relying on any individual EV data. 
All methods deliver similar frequency regulation performance. However, the eSSM requires both the collection and transmission of individual EV information, and the FL-based method still needs collecting such data locally. In contrast, the proposed approach operates solely on aggregated data, ensuring privacy and substantially reducing communication overhead.

\vspace{-0.4cm}
\subsection{The performance of Different Frequency Regulation Methods under Inaccurate SOC}
%As discussed before, in real applications, 
The EV SOC values and characteristic parameters can be incomplete or inaccurate for various reasons in practical applications. %In this section, we 
To test the robustness of different frequency regulation methods, we assume that SOC values reported to the aggregator may deviate by up to 10\% or 30\% from their true values. %different scenarios 
%when the EV parameters are inaccurate. %dataset gets inaccurate. 
%Specifically, we assume that the SOC values collected by the aggregator may deviate by up to 10\% \color{blue}/30\% \color{black}from their true values. 
To simulate this, we add additive Gaussian noise %is introduced to the SOC data, modeled as a normal distribution with zero mean and variance 2.0, i.e., 
$\epsilon \sim \mathcal{N}(0, 2.0)$ to the SOC data, truncated by $\pm 10$\% (moderate inaccuracy) and $\pm30$\% (severe inaccuracy) of the true values. 
\vspace{-0.3cm}
%where the noise $\epsilon$ is truncated within specified ranges to reflect different levels of uncertainty. 
%Two uncertainty levels are considered: $\epsilon \in [-0.1, 0.1]$ for moderate inaccuracy, and $\epsilon \in [-0.3, 0.3]$ for severe inaccuracy. 
%Corresponding distributions of power output errors are shown in Fig. \ref{fig:error_vari}.
\color{black}
\begin{table}[!htb]
\caption{Power Tracking Errors (MAE/MW) of Different Control Methods under SOC Inaccuracies}
\vspace{-0.2cm}
\label{tab:error_control_vari}
\centering
\renewcommand{\arraystretch}{1.2}
\begin{tabular*}{\linewidth}{@{\extracolsep{\fill}}lccc}
\hline\hline
SOC Inaccuracies & eSSM & FL-based & bHMM \\
\hline
$[-0.1, 0.1]$ & 0.1065 & 0.1813 & {0.0544} \\
$[-0.3, 0.3]$ & 0.1276 & 0.2403 & {0.0544} \\
\hline\hline
\end{tabular*}
% \vspace{-0.1cm}
\end{table}

\begin{figure}[!htb]
% \vspace{-0.5em}
    \centering
    \includegraphics[width=0.92\linewidth]{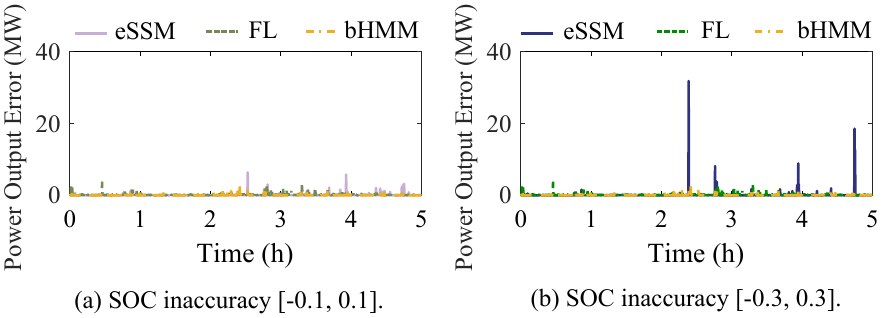}
    % \vspace{-0.3cm}
    \caption{Absolute power tracking error
    of different modeling methods under inaccurate SOC.}
    \label{fig:error_vari}
    % \vspace{-0.3cm}
\end{figure}
% \vspace{-0.5em}
\begin{figure}[!htb]
    \centering
    \includegraphics[width=0.92\linewidth]{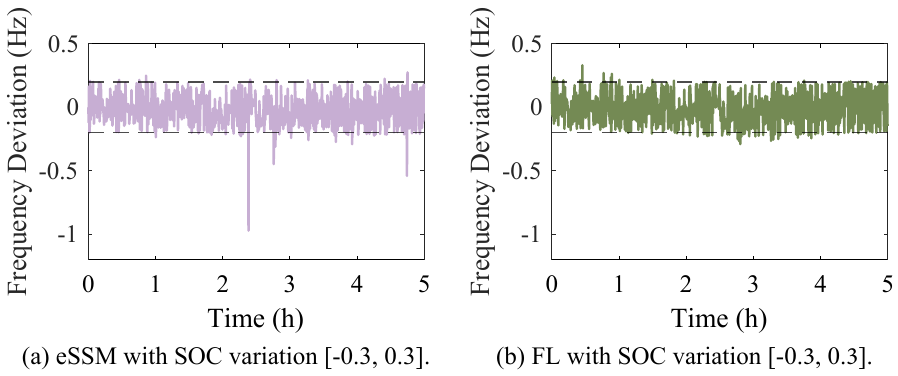}
    % \vspace{-0.3cm}
    \caption{Frequency deviation profiles of different modeling methods under inaccurate SOC.}
    \label{fig:frequency_soc}
    % \vspace{-0.5em}
\end{figure}

%\color{blue}In Scenario II, 
%The impact of such SOC data inaccuracies on power tracking errors of the eSSM and the FL-based method are shown in TABLE. \ref{tab:error_control_vari} and Fig. \ref{fig:error_vari}. %Power output errors and 
%The frequency regulation performance using eSSM and FL %different modeling approaches 
%is presented in %Fig.~\ref{fig:error_vari} %and  
%Fig.~\ref{fig:frequency_soc}. 
The impact of SOC data inaccuracies on the power-tracking performance of the eSSM and FL-based methods is summarized in Table \ref{tab:error_control_vari} and illustrated in Fig. \ref{fig:error_vari}, while their frequency regulation performance is shown in Fig. ~\ref{fig:frequency_soc}. Since the proposed method does not require any individual EV information, including SOC values, the resulting bHMM control performance under SOC inaccuracies remains identical to that shown in Fig.~\ref{fig:frequency_N2}. In contrast, SOC inaccuracies significantly degrade the performance of the eSSM and FL-based methods, which struggle to track the reference power $P^{\text{ref}}$. The proposed approach, however, demonstrates strong robustness to SOC errors, maintaining accurate power tracking and stable frequency regulation.
% In addition, the corresponding mean absolute errors (MAEs) are 0.057 MW (by the proposed method), 0.087 MW (by eSSM with 10\% inaccuracies), and 0.122 MW (by eSSM with 30\% inaccuracies).\color{blue}10\% FL: 7.9756\%; 30\% FL:\color{black}
%As shown in the figure, 
%These results show that SOC inaccuracies significantly degrade the 
%performance of the eSSM method and FL-based control method, which struggles to track the reference power $P^{ref}$. 
%In contrast, the proposed method demonstrates robustness against SOC inaccuracies, maintaining more accurate power tracking and more stable frequency regulation performance. \color{black}
% \vspace{-1em}

%and maintains more stable frequency regulation.

% To validate our model for missing data cases, we assume that 10\% of EVs have random degradation of capacity from 0\% to 30\%. Under this situation, simulation results are shown in Fig. {}. 

%---------------------------Conclusion-----------------------------
\section{Conclusion} %\color{red} TBD\color{black}
This paper proposed a data-driven privacy-preserving modeling and frequency regulation method using aggregated EVs. The proposed method accurately estimates the power output and flexibility of aggregated EVs and carries out effective frequency regulation without any individual EV information, addressing key privacy and data quality concerns. By leveraging a bHMM and EM algorithm, the framework maintains the scalability of model-based eSSM with minimal communication overhead, enabling real-time implementation. %The proposed method also has low communication overhead and can be implemented in real-time horizon. 
Simulation results show that the proposed method %also maintains the scalability of model-based eSSM while 
provides more effective frequency regulation (lower power tracking errors and smaller frequency deviations) than the eSSM and FL under SOC inaccuracies. In the future, we plan to consider incorporating user preferences into the framework %with different coefficients, \color{black} 
and extend the data-driven bHMM modeling approach to other distributed energy resources and ancillary services.    
\color{black}
% 2) Compared with the IMM and the SSM modeling, the data-driven model achieves high prediction accuracy.

% 3) The model realizes control over a large number of EVs with limited communication requirements by dispatching some authorities to individual EVs. It decreases the risk of cyber-attacks and transmission errors.

\bibliographystyle{IEEEtran}
% \bibliography{IEEEexample}
\bibliography{IEEEabrv,IEEEexample.bib}

%----------------------------Appendix-----------------------------

\appendices
\section{E Step: Kalman Filter and Smoother}
Let $\bm{V}_{0} = \bm{A} - \bm{I}$, and define $\bm{\tilde{V}}_{j}$ as in~\eqref{eq:Vj} for $j = 0, \ldots, N_{\bm{u}}$. The model in~\eqref{eq:bHMM} can be reformulated as~\eqref{eq:bls}.
\vspace{-0.1cm}
\begin{equation}
    {\footnotesize
        \bm{\tilde{V}}_{j} = \begin{bmatrix}
            0 & \bm{V}_{j}^{\prime} \\
            \bm{0} & \bm{V}_{j}^T
        \end{bmatrix}
        \label{eq:Vj}
    }
    \vspace{-0.1cm}
\end{equation}
\begin{equation}
\vspace{-0.5em}
{\footnotesize
    \begin{cases}
        \begin{bmatrix}
        1 \\
        \bm{x}_{k+1}
    \end{bmatrix} 
    = 
    \left( \bm{I} + \sum\limits_{j=0}^{N_{\bm{u}}}[u_{k}]_j\tilde{\mathbf{V}}_{j} \right)^{T}
    \begin{bmatrix} 1 \\ 
        \bm{x}_{k} 
    \end{bmatrix}  + \begin{bmatrix} 0 \\ 
    \bm{w}_{k}
    \end{bmatrix} 
    \\
    \bm{y}_{k} = \bm{c}_{0} + \bm{C}\bm{x}_{k} + \bm{v}_{k}
\end{cases}
}
\label{eq:bls}
\end{equation}
where $\bm{V}_{j}^{\prime}\in\mathbb{R}^{1\times (3N+3)}$ corresponds to the augmented state variable 1, and $\bm{\hat{V}}_{j} = [(\bm{V}_{j}^{\prime})^{T}\ \bm{V}_{j}]^{T}$. $\bm{c}_{0}$ is a constant offset to be estimated, and $k$ denotes the time step.

Define
\vspace{-0.1cm}
\begin{equation} \vspace{-0.5em} {\footnotesize \begin{bmatrix} 1 & \bm{b}_k^{T} \\ \bm{0} & \bm{A}_k^{T} \end{bmatrix} = \bm{I} + \sum\limits_{j=0}^{N_{\bm{u}}}[\bm{u}_{k}]_j\tilde{\bm{V}}_{j} } \end{equation}
then the bHMM dynamics can be expressed as a linear time-varying system:
\vspace{-0.2cm}
\begin{equation}
\vspace{-0.2cm}
{\footnotesize \begin{aligned} \bm{x}_{k+1} &= \bm{A}_{k}\bm{x}_{k}+\bm{b}_k+\bm{w}_{k}\\ \bm{y}_{k} &= \bm{c}_0 +\bm{C}\bm{x}_{k} + \bm{v}_k \end{aligned} } \label{eq:bilinear_equation} \end{equation}
In the expectation step, the model parameters $\bm{\Theta}$ is utilized to estimate the hidden states along each trajectory $\{\bm{y}_k\}_{k=0}^{K}$ and $\{\bm{u}_{k}\}_{k=0}^{K-1}$.  It includes the forward pass (Kalman filter) and backward recursion (smoother). 
During the forward pass, the Kalman filter computes the posterior covariance of each state $\bm{x}_k$ given the observations $\{\bm{y}_0, \bm{y}_1,\cdots,\bm{y}_k\}$. 

Denote the posterior means and covariances by 
\begin{equation}
    {\footnotesize
        \begin{aligned}
        \bm{\hat{\mu}_{n|k}} &= \mathbb{E}\left[\bm{x}_{n}|\bm{y}_{0},\cdots, \bm{y}_{k}\right] \\
    \bm{\hat{\Sigma}_{m,n|k}} &= \mathbb{E}\left[(\bm{x}_{n} - \bm{\hat{\mu}}_{m|k})(\bm{x}_{n} - \bm{\hat{\mu}}_{n|k})|\bm{y}_{0},\cdots,\bm{y}_{k}\right]
    \end{aligned}
    }
\end{equation}
Initially, since there are no observations from the trajectory, set 
\vspace{-0.2cm}
\begin{equation}
\vspace{-0.3em}
    {\footnotesize
        \bm{\hat{\mu}_{0|-1}} = \bm{\mu}_{0} \, \, \ \ \ \ \ 
    \bm{\hat{\Sigma}}_{0,0|-1} = \bm{\Sigma}_{0}
    }
\end{equation}
And given the previous conditional covariance $\bm{\hat{\Sigma}}_{k-1,k-1|k-1}$, the ``Kalman gain'' $\bm{K}_{k}$ is denoted as 
\vspace{-0.2cm}
\begin{equation}
\vspace{-0.4em}
    {\footnotesize
        \bm{K}_{k} = \bm{\hat{\Sigma}}_{k,k|k-1}\bm{C}^{T}\left(\bm{\Sigma_{v}} + \bm{C}\bm{\hat{\Sigma}}_{k,k|k-1}\bm{C}^{T}\right)^{-1}
    }
\end{equation}
So the next covariance and estimation of the current mean can be calculated by \eqref{eq:sig_kk} and \eqref{eq:mu_kk}.
\begin{equation}
\vspace{-0.4em}
    {\footnotesize
        \bm{\hat{\Sigma}}_{k,k|k} = \bm{\hat{\Sigma}}_{k,k|k-1} - \bm{K}_{k}\bm{C}\bm{\hat{\Sigma}}_{k,k|k-1}
    }
    \label{eq:sig_kk}
\end{equation}
\begin{equation}
\vspace{-0.4em}
    {\footnotesize
        \begin{cases}
        \bm{\hat{\mu}}_{k|k} = \bm{\hat{\mu}}_{k|k-1} + \bm{K}_{k}\left(\bm{y}_{k} -\bm{c}_0 - \bm{C}\bm{\hat{\mu}}_{k|k-1}\right) \\
        \bm{\hat{\mu}}_{k|k-1} = \bm{A}_{k-1}\bm{\hat{\mu}}_{k-1|k-1} + \bm{b}_{k-1}
    \end{cases}
    }
    \label{eq:mu_kk}
\end{equation}
Each $\bm{\hat{\mu}}_{k|k}$, $\bm{\hat{\Sigma}}_{k,k|k}$, $\bm{\hat{\Sigma}}_{k,k|k-1}$ are stored for the backward pass.

The backward pass refines the state estimate by incorporating the remaining observations ${\bm{y}_{k+1}, \ldots, \bm{y}_{K}}$ through an efficient recursive procedure. Initialized with the final forward estimate $\bm{\hat{\mu}}_{K|K}$ and the posterior mean $\bm{\hat{\mu}}_{k+1}$, the smoothing gain $\bm{J}_{k}$ is given by~\eqref{eq:J_k}, and the smoothed mean and covariance at time step $k$ are computed as \eqref{eq:mean}-\eqref{eq:variance}.
\vspace{-0.2cm}
\begin{equation}
    {\footnotesize
        \bm{J}_{k} = \bm{\hat{\Sigma}}_{k,k|k}\bm{A}_{k}^{T}\left(\bm{\hat{\Sigma}}_{k+1,k+1|k}\right)^{-1}
    }
    \label{eq:J_k}
    \vspace{-0.1cm}
\end{equation}
\begin{equation}
\vspace{-0.4em}
\label{eq:mean}
    {\footnotesize
        \bm{\hat{\mu}}_{k} = \bm{\mu}_{k|k} + \bm{J}_{k}\left(\bm{\hat{\mu}}_{k+1} - \bm{A}_{k}\bm{\hat{\mu}}_{k|k} - \bm{b}_{k}\right)
    }
\end{equation}
\begin{equation}
\vspace{-0.4em}
\label{eq:variance}
    {\footnotesize
        \begin{cases}
    \bm{\hat{\Sigma}}_{k,k+1} = \bm{J}_{k}\bm{\hat{\Sigma}}_{k+1,k+1}\\
        \bm{\hat{\Sigma}}_{k,k} = \bm{\hat{\Sigma}}_{k,k|k} + \bm{J}_{k}\left(\bm{\hat{\Sigma}}_{k+1,k+1} - \bm{\hat{\Sigma}}_{k+1,k+1|k}\right)\bm{J}_{k}^{T} 
    \end{cases}
    }
\end{equation}

Upon completing the forward–backward passes, the inference distribution parameters $\{\bm{\hat{\mu}}_{k}\}_{k=0}^{K}$, $\{\bm{\hat{\Sigma}}_{k,k}\}_{k=0}^{K}$, and $\{\bm{\hat{\Sigma}}_{k,k+1}\}_{k=0}^{K-1}$ are available for the maximization step.

\section{M Step: Analytical Solution Derivation}

Define the mean and joint covariance of the inference distribution $Q^{l},l=1,\cdots, L$ as \eqref{eq:def_mu}.
\vspace{-0.1cm}
\begin{equation}
{\footnotesize
    \hat{\bm{\mu}}_{j}^{l}= \mathbb{E}_{\bm{\hat{X}}^{l}}\left[\bm{\hat{x}}_{j}^{l}\right] \, \, \ \ \ \ \bm{\hat{\Sigma}}_{j,k}^{l} = \mathbb{E}_{\bm{\hat{X}}^{l}}\left(||\bm{\hat{x}}^{l}_{j}-\bm{\hat{\mu}}_{j}^{l}||^{2}\right)
    \label{eq:def_mu}
    }
\end{equation}

%\color{red} 
As proved in  \cite{otto2022learning}\color{black}, with $\bm{C}$ fixed, the optimal parameters $\bm{\Theta}$ can be calculated analytically by following equations:
\vspace{-0.2cm}
\begin{equation}
\vspace{-0.4em}
    {\footnotesize
        \bm{\mu}_{0} = \frac{1}{L}\sum_{l=1}^{L}\bm{\hat{\mu}}_{0}^{l}, \, \, \ \ \bm{\Sigma}_{0} = \frac{1}{L}\sum_{l=1}^{L}\left(\bm{\hat{\Sigma}}_{0,0}^{l} + ||\bm{\hat{\mu}}_{0}^{l} - \bm{\mu}_{0}||^{2}\right)
    }
    \label{eq:optimal_mu}
\end{equation}
%     \begin{equation}
%     \vspace{-0.4em}
%         {\footnotesize
            
%         }
%         \label{eq:cal_sigma_0}
%     \end{equation}
\begin{equation}
\vspace{-0.4em}
    {\footnotesize
        \bm{c}_{0} = \frac{1}{L(K+1)}\sum\limits_{l=1}^{L}\sum\limits_{k=0}^{K}\left(\bm{y}_{k}^{l} - \bm{C}\bm{\hat{\mu}}_{k}^{l}\right)
     }
\end{equation}
    \begin{equation}
    \vspace{-0.4em}
        {\footnotesize
            \bm{\Sigma}_{\bm{v}}  = \frac{1}{L\left(K+1\right)}\sum_{l=1}^{L}\sum_{k=0}^{K}\bigg(\bm{C}\bm{\hat{\Sigma}}_{k,k}^{l}\bm{C}^{T} + ||\bm{y}_{k}^{l}- \bm{c}_0 - \bm{C}\bm{\hat{\mu}}_{k}^{l}||^{2}\bigg)
        }
    \end{equation}
\begin{equation}
\vspace{-0.4em}
        {\footnotesize
            \begin{aligned}
            \bm{\Sigma}_{\bm{w}}  = &\frac{1}{KL}\sum_{l=1}^{L}\sum_{k=0}^{K-1}\bigg[\bm{\Sigma}_{k+1,k+1}^{l} - \bm{A}_{k}^{l}\bm{\hat{\Sigma}}_{k,k+1}^{l} - \bm{\hat{\Sigma}}_{k+1,k}^{l}(\bm{A}_{k}^{l})^{T} \\&+ \bm{A}_{k}^{l}\bm{\hat{\Sigma}}_{k,k}^{l}(\bm{A}_{k}^{l})^{T} + ||\bm{\hat{\mu}}_{k+1}^{l} - \bm{A}_{k}^{l}\bm{\hat{\mu}}_{k}^{l} - \bm{b}_{k}^{l}||^2\bigg]
        \end{aligned}
        }
    \end{equation}
\begin{equation}
% \vspace{-0.4em}
\label{eq:cal_V}
    {\footnotesize
        \begin{aligned}
        \begin{bmatrix}
        \bm{\hat{V}}_{0} \\ \vdots \\ \bm{\hat{V}}_{N_{\bm{u}}}
    \end{bmatrix} = 
    \left( \sum\limits_{l=1}^{L} \sum\limits_{k=0}^{K-1}\bm{u}_{k}^{l} \otimes (\bm{u}_{k}^{l})^{T} \otimes \bm{G}_{k}^{l} \right)^{-1}
    \left(\sum\limits_{l=1}^{L} \sum\limits_{k=0}^{K-1} \bm{u}_{k}^{l} \otimes (\bm{\tilde{H}}_{k}^{l})^{T}\right)
    \end{aligned}
    }
\end{equation}
where $\otimes$ denotes the Kronecker product, and $\bm{G}_{k}^{l}$ and $\bm{\tilde{H}}_{k}^{l}$ are given by \eqref{eq:cal_Gl} %\color{red} check \color{black} 
and \eqref{eq:cal_Hl}.
\begin{equation}
\vspace{-0.4em}
    {\footnotesize
        \begin{aligned}
            \bm{G}_{k}^{l} = \begin{bmatrix}
                1 & \left(\bm{\hat{\mu}}_{k}^{l}\right)^{T} \\
                \bm{\hat{\mu}}_{k}^{l} & \bm{\hat{\Sigma}}_{k,k}^{l} + \bm{\hat{\mu}}_{l}^{\left(m\right)}\left(\bm{\hat{\mu}}_{k}^{l}\right)^{T}
            \end{bmatrix}
        \end{aligned}
    }
    \label{eq:cal_Gl}
\end{equation}
\begin{equation}
\vspace{-0.4em}
    {\footnotesize
        \begin{aligned}
            \bm{\tilde{H}}_{k}^{l} = \begin{bmatrix}
                \bm{\hat{\mu}}_{k+1}^{l} - \bm{\hat{\mu}}_{k}^{l} \\ \bm{\hat{\Sigma}}_{k+1,k}^{l} - \bm{\hat{\Sigma}}_{k,k}^{l} + \left(\bm{\hat{\mu}}_{k+1}^{l} - \bm{\hat{\mu}}_{k}^{l}\right)\left(\bm{\hat{\mu}}_{k}^{l}\right)^{T}
            \end{bmatrix} ^{T}
        \end{aligned}
    }
    \label{eq:cal_Hl}
\end{equation}

The above solution exists and is unique provided that the empirical information matrix
\begin{equation}
    \sum\limits_{l=1}^{L} \sum\limits_{k=0}^{K-1}\bm{u}_{k}^{l} \otimes (\bm{u}_{k}^{l})^{T} \otimes \bm{G}_{k}^{l}
\end{equation}
is nonsingular. This condition corresponds to the classical persistent excitation requirement in system identification theory, i.e., the joint regressor must yield a full-rank covariance matrix to ensure identifiability of the bilinear parameters. To verify this condition in practice, we evaluated the minimum eigenvalue of the normalized empirical information matrix during EM iterations. Across all training runs, the smallest eigenvalues remain strictly positive (greater than $10^{-8}$), indicating that the regressor covariance matrix is numerically full rank and that the persistent excitation condition is satisfied by the dataset.

\end{document}